%% file: mainsm.tex
\documentclass[aps, prl, twocolumn, tightenlines, letterpaper, amsmath, amssymb, preprintnumbers, superscriptaddress, floatfix, longbibliography, nofootinbib]{revtex4-2}
\usepackage[T1]{fontenc}
\usepackage{bm}
\usepackage{xspace}
\usepackage[dvipsnames]{xcolor}
\usepackage{graphicx}
\usepackage{tabularx}
\usepackage{enumitem}
\usepackage{qcircuit}

\usepackage[hypertexnames=false]{hyperref}
\hypersetup{
    colorlinks=true,       
    linkcolor=blue,          
    citecolor=blue,        
    filecolor=blue,      
    urlcolor=blue           
}

\newcolumntype{Y}{>{\centering\arraybackslash}X}
\newcolumntype{L}{>{\raggedright\arraybackslash}X}

\usepackage{mathtools}
\usepackage{orcidlink}
\usepackage[normalem]{ulem}

\newcommand{\beginsupplement}{%
        \renewcommand{\thefigure}{S\arabic{figure}}
        \renewcommand{\theHfigure}{S\arabic{figure}}
        \setcounter{figure}{0}
        \renewcommand{\thetable}{S\arabic{table}}
        \renewcommand{\theHtable}{S\arabic{table}}
        \setcounter{table}{0}
        \renewcommand{\theequation}{S\arabic{equation}}
        \renewcommand{\theHequation}{S\arabic{equation}}
        \setcounter{equation}{0}
        \renewcommand{\thesection}{S\arabic{section}}
        \renewcommand{\theHsection}{S\arabic{section}}
        \setcounter{section}{0}
     }

\begin{document}

\title{Multi-Neutrino Entanglement and Correlations in Dense Neutrino Systems
}

\author{Marc~Illa\,\orcidlink{0000-0003-3570-2849}}
\email{marcilla@uw.edu}
\affiliation{InQubator for Quantum Simulation (IQuS), Department of Physics, University of Washington, Seattle, WA 98195}
\author{Martin J.~Savage\,\orcidlink{0000-0001-6502-7106}}
\email{mjs5@uw.edu}
\affiliation{InQubator for Quantum Simulation (IQuS), Department of Physics, University of Washington, Seattle, WA 98195}

\preprint{IQuS@UW-21-034}
\date{\today}

\begin{abstract} 
The time-evolution of multi-neutrino entanglement and correlations are 
studied in two-flavor collective neutrino oscillations, relevant for dense neutrino environments, building upon previous works. 
Specifically, simulations performed of systems with up to 12 neutrinos using Quantinuum's \texttt{H1-1} 20 qubit trapped-ion quantum computer are used to compute
n-tangles, and  two- and three-body correlations, probing beyond mean-field descriptions.
$n$-tangle re-scalings are found to converge for large system sizes, signaling the presence of genuine multi-neutrino entanglement. 
\end{abstract}

\maketitle


In extreme astrophysical environments, such as those found in core-collapse supernovae,
neutrino densities are sufficiently high to participate in the 
transport of energy and momentum, in local chemical compositions and in dynamics~\cite{Duan:2009cd,Duan:2010bg,Chakraborty:2016yeg,Tamborra:2020cul,Capozzi:2022slf}.
Coherent evolution of lepton flavors, that depends on
self-interactions between neutrinos induced by weak interactions~\cite{1987ApJ...322..795F,Savage:1990by,PhysRevD.46.510,PANTALEONE1992128,Bruce_PhysRevD.49.2710},  plays an important role. 
First studies of the quantum correlations in coherent evolution of dense neutrino systems, beyond mean-field descriptions, 
are providing important insights into such dynamics~\cite{Bell:2003mg,Friedland:2003eh,Sawyer:2004ai,Pehlivan:2011hp,Rrapaj:2019pxz,Cervia:2019res,Roggero:2021asb,Hall:2021rbv,Roggero:2021fyo,Yeter-Aydeniz:2021olz,Patwardhan:2021rej,Xiong:2021evk,Martin:2021bri,Cervia:2022pro,Illa:2022jqb,Roggero:2022hpy,Lacroix:2022krq,Amitrano:2022yyn}.
So far, they have focused on bipartite entanglement witnesses, such as entanglement entropy, negativity and concurrence~\cite{Cervia:2019res,Rrapaj:2019pxz,Roggero:2021asb,Hall:2021rbv,Roggero:2021fyo,Patwardhan:2021rej,Xiong:2021evk,Martin:2021bri,Cervia:2022pro,Illa:2022jqb,Roggero:2022hpy}.
In this work, 
we explore multi-neutrino entanglement in such systems 
by computing 
$n$-tangles~\cite{PhysRevA.63.044301}, $\tau_n$, between $n$ neutrinos
induced by time evolution.
The total $n$-tangles at late times are found to scale for large system sizes.
Our work utilized classical simulations, and quantum simulations
using the Quantinuum 20-qubit trapped-ion quantum computer {\tt H1-1} and noisy emulator {\tt H1-1E}~\cite{QuantHoney}.

The leading-order low-energy effective Hamiltonian describing collective, coherent 
neutrino flavor oscillations is composed of three terms.
One term is responsible for vacuum oscillations, originating from the neutrino mass matrix~\cite{Pontecorvo:1957cp,Pontecorvo:1957qd,Maki:1962mu,Pontecorvo:1967fh}.
A second is from the weak 
interactions between neutrinos and matter, mainly between $\nu_e$ and $e^-$,  through charged-current processes, and is responsible for the 
Mikheev-Smirnov-Wolfenstein effect~\cite{Wolfenstein:1977ue,Mikheyev:1985zog}.
In what follows, we neglect the contributions from this term.
A third term, from the neutral-current weak interactions, 
is responsible for coherent forward scattering of neutrinos,
which becomes significant at sufficiently high neutrino densities~\cite{Savage:1990by,PhysRevD.46.510,PANTALEONE1992128,Bruce_PhysRevD.49.2710}.

Due to the small value of $\theta_{13}$~\cite{Esteban:2020cvm}, 
three-flavor neutrino systems 
can be approximated by two-flavor systems involving 
the electron neutrino $\nu_e$ and a heavy neutrino $\nu_x$, 
considered to be a combination of 
$\nu_\mu$ and $\nu_\tau$~\cite{Balantekin:1999dx}.
The effective Hamiltonian for $N$ neutrinos can be written in terms of 
spin operators acting in flavor space~\cite{Pehlivan:2011hp},
\begin{equation}
    H=H^{\nu}+H^{\nu\nu}=\sum_i \mathbf{b}\cdot \bm{\sigma}^{(i)}+\frac{1}{N} \sum_{i<j} J_{ij}\bm{\sigma}^{(i)}\cdot \bm{\sigma}^{(j)} \ ,
    \label{eq:ham2flavor}
\end{equation}
where $\bm{\sigma}^{(i)}=(\sigma^{(i)}_x, \sigma^{(i)}_y, \sigma^{(i)}_z)$ 
are the Pauli matrices acting on the $i^{\rm th}$ neutrino flavor-doublet, 
$\left(|\nu_e\rangle , |\nu_x\rangle \right)^T$.
The 3-vector $\mathbf{b}$ encodes the vacuum oscillations,
\begin{equation}
    \mathbf{b} = \frac{\Delta m^2}{4E}
    (\sin(2\theta_v),0,-\cos(2\theta_v)) 
    \ ,
    \label{eq:bdef}
\end{equation}
with $\Delta m^2$ being the difference between neutrino squared-masses, 
$E$ the neutrino energy, 
and $\theta_v$ the vacuum-mixing angle.
The two-body couplings $J_{ij}$, 
uniquely defined at leading order by the Standard Model,
are,
\begin{equation}
    J_{ij} = \sqrt{2}G_F\rho_{\nu}(1-\cos\theta_{ij}) \ ,
\end{equation}
where $G_F$ is Fermi's constant, 
$\rho_{\nu}$ the number-density of neutrinos,
and $\theta_{ij}$ the angle of the momenta between the $i^{\rm th}$ and $j^{\rm th}$ neutrino.
Following previous works, e.g., Refs.~\cite{Hall:2021rbv,Amitrano:2022yyn}, 
we introduce $\mu\equiv \sqrt{2}G_F \rho_{\nu}$, 
and,
for demonstrative purposes,
set $\mu/N=\Delta m^2/4E$ (assuming a monochromatic beam), 
so that the one- and two-body terms have comparable strengths~\cite{Hall:2021rbv}.
Following the inspiring work of Hall {\it et al.}~\cite{Hall:2021rbv}, 
a one-parameter set is used to demonstrate relevant physics, with a 
vacuum-mixing angle of $\theta_v=0.195$, 
a distribution of momenta with
$\theta_{ij}=\arccos(0.9)\times|i-j|/(N-1)$ (cone-shaped), 
and an initial state that is a product state of $N/2$ $|\nu_e\rangle$ and $N/2$ $|\nu_x\rangle$, 
i.e., 
$|\Psi_0\rangle=|\nu_e\rangle^{\otimes N/2}\otimes|\nu_x\rangle^{\otimes N/2}$.

While a number of previous calculations of neutrino systems 
using quantum devices have been focused on the coherent time evolution of the flavor content~\cite{Hall:2021rbv,Yeter-Aydeniz:2021olz,Illa:2022jqb,Amitrano:2022yyn}, 
and the entanglement of one or two neutrinos~\cite{Hall:2021rbv,Illa:2022jqb}, 
here we examine correlations between different neutrinos 
and multi-body entanglement, 
quantities that can further probe  mean-field descriptions of these systems.
Specifically, we look at the two-body correlations, as considered previously, e.g., Ref.~\cite{Roggero:2022hpy},
\begin{equation}
    \mathcal{C}^{(2\nu)}_{ij}(t)=\langle \sigma^{(i)}_z \sigma^{(j)}_z\rangle - \langle \sigma^{(i)}_z \rangle \langle\sigma^{(j)}_z\rangle\ ,\quad i\neq j \ ,
\end{equation}
with $\langle \cdot \rangle =\langle \Psi_t |\cdot| \Psi_t \rangle$,
and  three-body correlations,
\begin{align}
    \mathcal{C}^{(3\nu)}_{ijk}(t)=&\ \langle \sigma^{(i)}_z \sigma^{(j)}_z\sigma^{(k)}_z\rangle - \langle \sigma^{(i)}_z \rangle \langle\sigma^{(j)}_z\sigma^{(k)}_z\rangle \nonumber\\
    &- \langle \sigma^{(j)}_z \rangle \langle\sigma^{(i)}_z\sigma^{(k)}_z\rangle- \langle \sigma^{(k)}_z \rangle \langle\sigma^{(i)}_z\sigma^{(j)}_z\rangle \nonumber\\
    &+2 \langle \sigma^{(i)}_z \rangle \langle\sigma^{(j)}_z\rangle\langle\sigma^{(k)}_z\rangle\ ,\quad i\neq j \neq k \ .
\end{align}
Ultimately, studies of entanglement will help quantify 
the limitations of classical techniques in providing reliable 
and accurate results, and can also 
guide tensor-networks approaches~\cite{Roggero:2021asb,Roggero:2021fyo,Cervia:2022pro}.
Entanglement in neutrino systems has been previously computed via full-state tomography~\cite{Hall:2021rbv,Illa:2022jqb}, 
which, however, is expected to become inefficient for larger systems 
due to the required number of measurements.
Alternative methods including classical shadows~\cite{Huang_2020}, 
which are expected to require fewer measurements, 
are currently being pursued.

As a way to gain further insight into the entanglement structure of dense, coherent neutrino systems, 
we focus on the $n$-tangle $\tau_n$, 
defined as $|\langle \Psi|\sigma^{\otimes n}_y|\Psi^*\rangle |^2$,
where $|\Psi^*\rangle$ is the complex conjugate of $|\Psi\rangle$. 
This quantity is a measure of the $n$-body entanglement~\cite{PhysRevA.63.044301}. 
For two-qubit (neutrino) systems, $\tau_2$ is the concurrence squared~\cite{Hill:1997pfa,Wootters:1997id}, $C^2_{12}$.
For three-qubit systems, 
$\tau_3$ is the residual entanglement~\cite{Coffman:1999jd}, $C^2_{1(23)}-C^2_{12}-C^2_{13}$, with $C_{1(23)}=\sqrt{2-2\text{Tr}\rho^2_1}$, 
with $\rho_1$ the reduced density matrix of the first qubit.
However, for $n\geq 4$, $\tau_n$ is not uniquely the residual entanglement, 
and it is smaller than or equal to the concurrence 
$C_{1(2\ldots n)}$~\cite{Li:1003.4774}. 
An interesting property of 
$\tau_n$ is that, 
for the $N$-qubit GHZ state~\cite{greenberger1989going,Bouwmeester:1998iz}, $\tau_N=1$, 
while for the $N$-qubit W state~\cite{Dur:2000zz}, $\tau_N=0$. 
Thus, while it can help distinguish between different types of entanglement,
by itself it is not a measure of $N$-body entanglement~\cite{PhysRevA.63.044301}.\footnote{
There are other ways to ascertain the nature of entanglement in these systems. 
For example, a more complete set of two-point correlation functions could be used to determine the Quantum Fisher information, $F=\sum_{\alpha,\beta}\sum_{i,j}n_{\alpha}^iC_{\alpha,\beta}^{i,j}n_{\beta}^j$, with $n_{\alpha}^i$ being unit vectors and $C_{\alpha,\beta}^{i,j}=\langle \sigma^{(i)}_\alpha\sigma^{(j)}_{\beta}\rangle -\langle \sigma^{(i)}_\alpha\rangle \langle \sigma^{(j)}_{\beta}\rangle$, 
which can be used to learn about the nature of entanglement~\cite{2012PhRvA..85b2321H,2012PhRvA..85b2322T}.
}


\textit{Implementation on a Quantum Computer ---} 
Previous works have performed quantum simulations
of these systems using IBM's superconducting-qubit quantum computers~\cite{Hall:2021rbv,Yeter-Aydeniz:2021olz},
D-Wave's superconducting-qubit annealing devices~\cite{Illa:2022jqb}
and Quantinuum's Trapped-ion quantum computers~\cite{Amitrano:2022yyn}.
We implement the time evolution of an initial state $\Psi_0$ 
under the Hamiltonian given in Eq.~\eqref{eq:ham2flavor} 
using Quantinuum's {\tt H1-1} 20 qubit trapped-ion quantum computer, 
that is based on using the two hyperfine clock states in the ${}^2S_{1/2}$ ground state of ${}^{172}$Yb${}^+$ ions as qubits~\cite{QuantHoney}. 
For  purposes of comparison, we also provide results obtained using its 
emulator, {\tt H1-1E}.\footnote{
Ref.~\cite{Farrell:2022vyh} showed that  {\tt H1-1E} could well reproduce the behavior of 
{\tt H1-1}.}
The evolution operator, $\exp(-itH)$, is Trotterized, and each contribution is mapped to a quantum circuit composed of gates that are native to {\tt H1-1}.
As noted in Ref.~\cite{Amitrano:2022yyn}, 
the one-body and two-body parts of the Hamiltonian commute, 
$[H^{\nu},H^{\nu\nu}]=0$, 
so they can be Trotterized without introducing a higher-order 
systematic error. 
For $H^{\nu}$, since each term acts on a different neutrino,
\begin{equation}
    e^{-it H^{\nu}}=\bigotimes_{i}e^{-it \mathbf{b}\cdot \bm{\sigma}^{(i)}} \ .
\end{equation}
This term is implemented with the following Euler decomposition for $SU(2)$ matrices~\cite{Barenco:1995na},
\begin{equation}
e^{-it \mathbf{b}\cdot \bm{\sigma}^{(i)}} =\ \begin{gathered}
     \Qcircuit @R 0.7em @C 1em {
     & \gate{R_z(\alpha_1)} & \gate{R_y(\alpha_2)} & \gate{R_z(\alpha_3)} & \qw 
     }
\end{gathered}\ ,
\label{eq:1bodyop}
\end{equation}
where the angles $\alpha_i$ are fixed numerically.

For $H^{\nu\nu}$,  the known decomposition of $SU(4)$ matrices with 3 CNOT gates~\cite{vatan2004optimal} is used, 
\begin{align}
&e^{-it J_{ij}\bm{\sigma}^{(i)}\cdot \bm{\sigma}^{(j)}} = \nonumber \\
&\begin{gathered}
    \resizebox{.88\hsize}{!}{\Qcircuit @R 0.7em @C 0.5em {
    & \qw & \targ & \gate{R_z(\beta-\frac{\pi}{2})} & \ctrl{1} & \qw & \targ & \gate{R_z(-\frac{\pi}{2})} & \qw \\
    & \gate{R_z(\frac{\pi}{2})} & \ctrl{-1} & \gate{R_y(\frac{\pi}{2}-\beta)} & \targ & \gate{R_z(\beta-\frac{\pi}{2})}& \ctrl{-1} & \qw &\qw
    }}
\end{gathered}\ ,
\label{eq:2bodyop}
\end{align}
with $\beta=2tJ_{ij}$, 
which has a slight advantage over~\cite{PhysRevA.69.010301,PhysRevA.77.066301} (requiring 5 single-qubits gates 
instead of 8). 
While not directly relevant to the present set of simulation, it is interesting to consider the T-gate resource requirements for such simulation.
Standard methods\footnote{Following Ref.~\cite{2013arXiv1311.1074P}, 
and assuming no additional ancilla qubits are used, 
for general single-qubit rotations $U_1$, Refs.~\cite{2004quant.ph.11206F,2012PhRvL.109s0501B} give the following estimates of the number of $T$-gates, $T_{U1}(\epsilon)=2.95\log_2(1/\epsilon) + 3.75$.
For $R_z$, Ref.~\cite{2012arXiv1212.6964K} gives $T_{Rz}(\epsilon)=3.21 \log_2(1/\epsilon) - 6.93$.
} suggest a T-gate count of
$N(187N-101)/2$ for  $\epsilon=10^{-4}$, for $N$ neutrinos per Trotter step.
While the circuits in Eqs.~\eqref{eq:1bodyop} and~\eqref{eq:2bodyop} are not written in terms of the native gates used in \texttt{H1-1} (these can be found in Ref.~\cite{Amitrano:2022yyn}), the package \texttt{pytket}~\cite{sivarajah2020t} includes a function that performs this translation, along with optimizations. 
\begin{figure}[!tb]
	\includegraphics[width=\columnwidth]{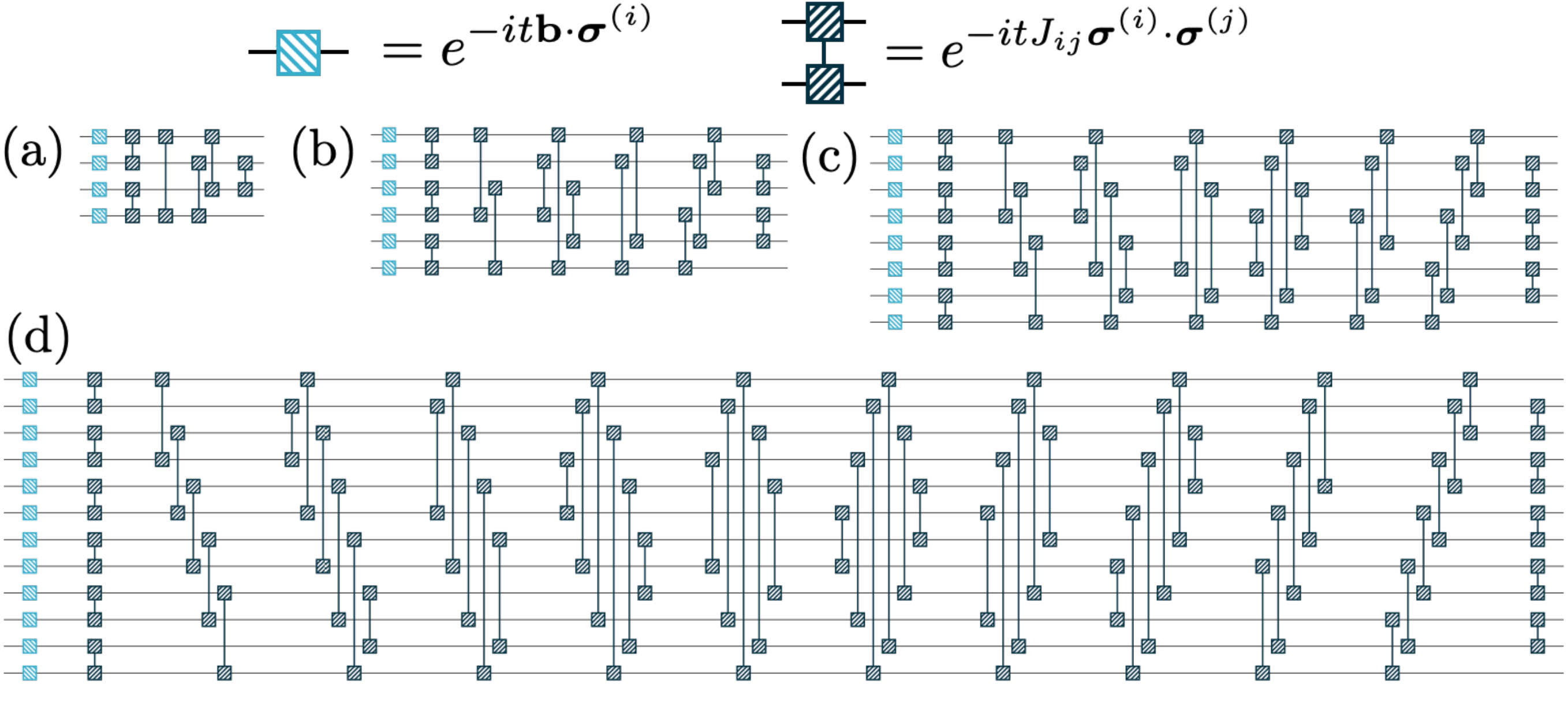}
	\caption{Circuits for a single Trotter step for (a) $N=4$, (b) $N=6$, (c) $N=8$, and (d) $N=12$ neutrinos, where the two-neutrino gates are clustered into groups that can be applied in parallel, showing the linear scaling with $N$ of the circuit depth.}
	\label{fig:circuits}
\end{figure}
Implementation of this two-neutrino term is more delicate. 
Since the sum over different pairs of neutrinos is split 
in the implementation of the time-evolution operator, 
the non-commutativity of terms introduces systematic Trotter errors.
While it is possible to find a combination that minimizes this error, as in Ref.~\cite{Amitrano:2022yyn} for the case of 4 neutrinos, this is not feasible for larger systems.
As shown in Ref.~\cite{Hall:2021rbv}, it is possible to build a circuit that performs the Trotterized version of $\exp(-itH^{\nu\nu})$ with $N$ layers of the operator in Eq.~\eqref{eq:2bodyop} (if multiple gates can be applied in parallel across the device).
Figure~\ref{fig:circuits} shows the circuits used for different number of neutrinos.
A nice property of these circuits is that they retain the symmetry 
present in the Hamiltonian given in Eq.~\eqref{eq:ham2flavor}, 
between the exchange of the $i^{\rm th}$ and $(N-i+1)^{\rm th}$ neutrino, 
for the current choice of $J_{ij}$. 
While first order Trotter evolution has been used, 
higher orders with their improved convergence have been explored in Ref.~\cite{Amitrano:2022yyn}. 
They found that the Trotter errors from first and second order evolution from the prepared initial state
are significantly smaller than naive theoretical bounds.
Alternative methods for time evolution, such as Variational Fast Forwarding~\cite{2020npjQI...6...82C}, 
should also be examined.


\textit{Two- and Three-Neutrino Correlations ---} 
A term that appears in the computation of correlation functions is the expectation value of $\sigma_z$. 
For this reason, it is interesting to 
first compute the inversion probability for a single neutrino, defined as
\begin{equation}
    P_i(t)=\frac{1}{2}(1\mp \langle \sigma_z^{(i)} \rangle) \ ,
\end{equation}
where the $\mp$ sign depends on the initial state of the $i^{\rm th}$ 
neutrino ($-$ for $\nu_e$ and $+$ for $\nu_x$).
Due to the symmetry of the Hamiltonian (and its Trotterized version), 
the inversion probabilities for the $i^{\rm th}$ 
and $(N-i+1)^{\rm th}$ neutrino are the same.
\begin{figure}[!tb]
	\includegraphics[width=\columnwidth]{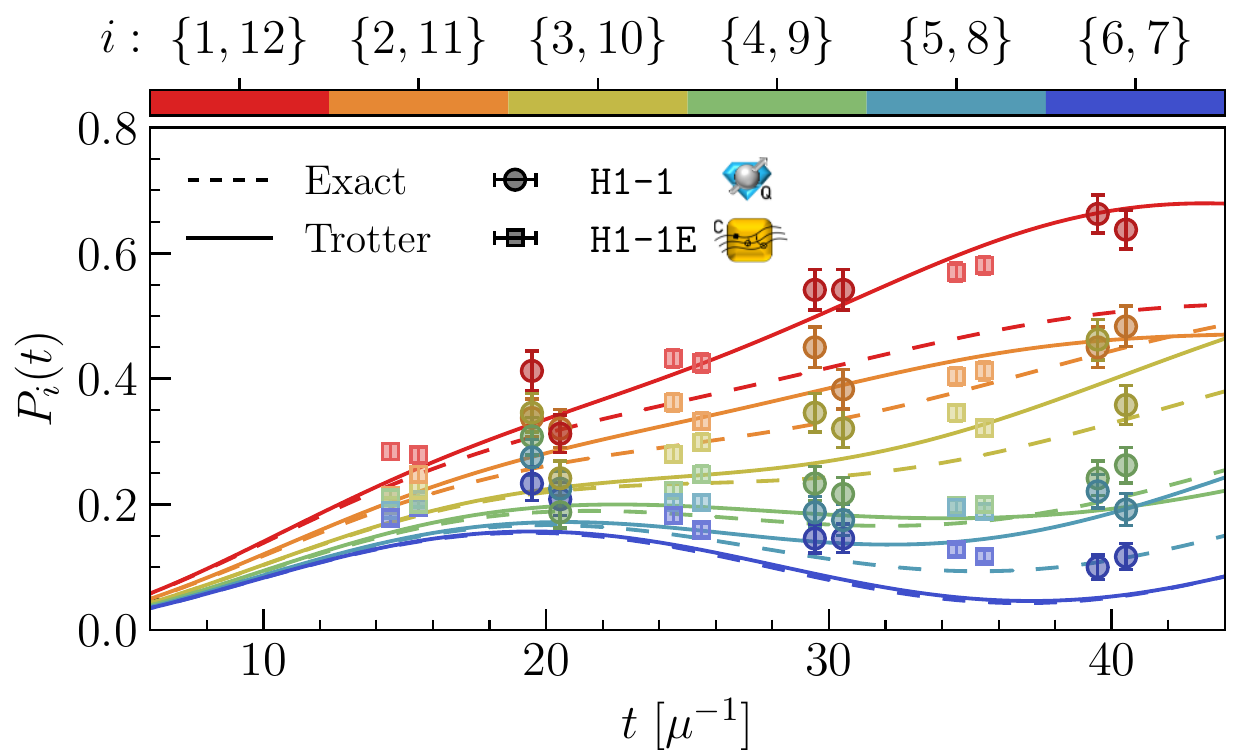}
	\caption{Flavor inversion probabilities for $N=12$ neutrinos. 
	The lines show the single-step Trotter (continuous) and exact (dashed) simulations, 
	and the points show the results from {\tt H1-1} (dark circles, using 240 shots) 
	and {\tt H1-1E} (light squares, using 1200 shots).}
	\label{fig:time_ev}
\end{figure}
The results for $N=12$ are shown in Fig.~\ref{fig:time_ev}, 
where the two icons in the legend indicate whether a quantum device (blue) or emulator (yellow) was used to obtain the results~\cite{Klco:2019xro}, and the $N=8$ are in the Supplemental Material (SM). 
The uncertainties in the results are computed via bootstrap resampling. 
For the $N=8$ case, agreement is found with the results presented in Ref.~\cite{Amitrano:2022yyn}. 
For $N=12$, both results from {\tt H1-1} and {\tt H1-1E} are showing somewhat larger deviations from expectations,  but the device 
has remained coherent.\footnote{The associated SM gives the numerical values of the results shown in Fig.~\ref{fig:time_ev}, and those shown in subsequent figures.}

As a selection of two- and three-body correlations, we focus on
\begin{itemize}[leftmargin=*]
\item[]
\renewcommand{\arraystretch}{1.5}
\begin{tabularx}{0.96\columnwidth}{L l} 
$\mathcal{C}^{(2\nu)}_{1i}$\ , with $1<i\leq N$\ , & $\mathcal{C}^{(3\nu)}_{1iN}$\ , with $1<i<N$\ , \\
$\mathcal{C}^{(2\nu)}_{i,i+1}$\ , with $1\leq i<N$\ , & $\mathcal{C}^{(3\nu)}_{i-1,i,i+1}$\ , with $1<i<N$\ . \\ 
\end{tabularx}
\renewcommand{\arraystretch}{1}
\end{itemize}
\begin{figure}[!tb]
	\includegraphics[width=\columnwidth]{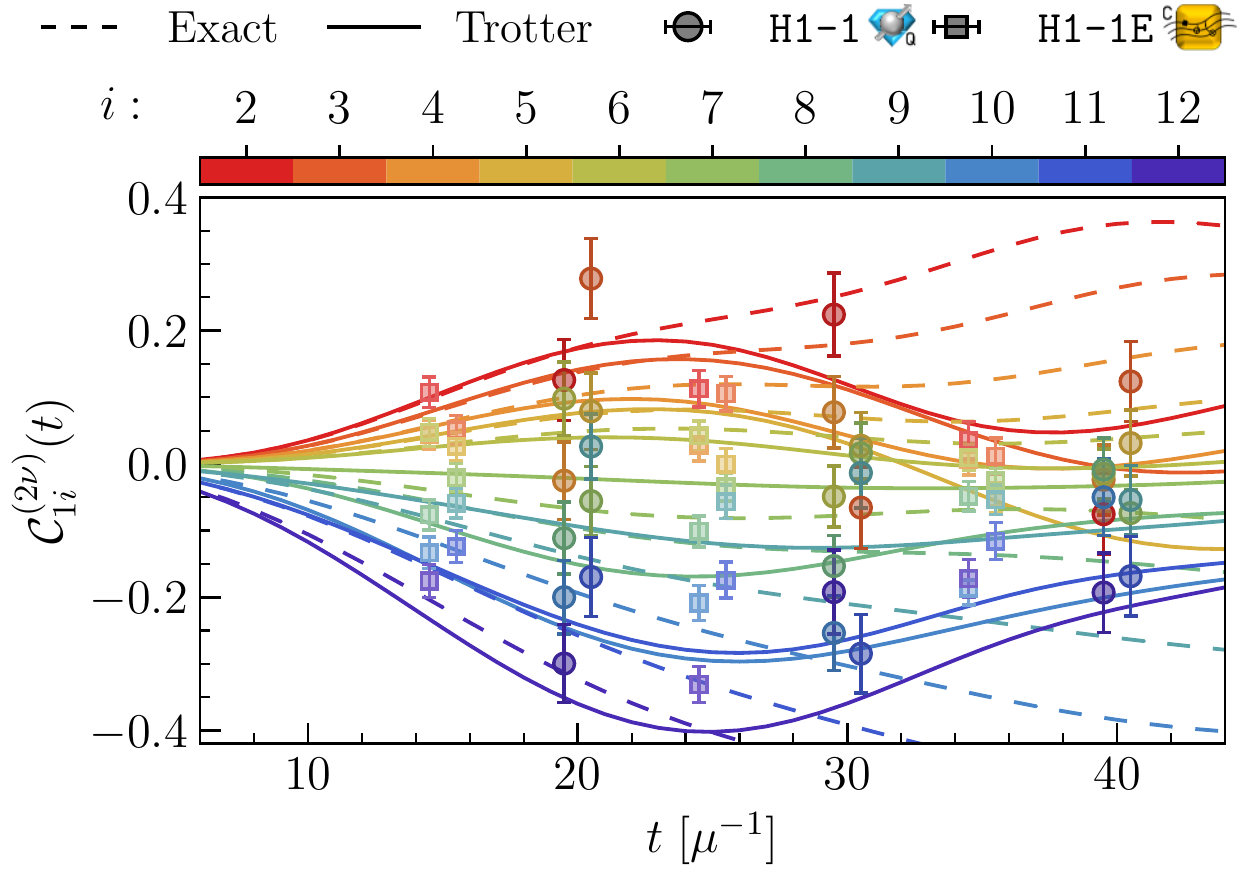}
	\caption{Two-body correlation for $N=12$ neutrinos. The lines show the single-step Trotter (continuous) or exact (dashed) simulation, and the points show the results from {\tt H1-1} (dark circles, using 240 shots) and {\tt H1-1E} (light squares, using 1200 shots).}
	\label{fig:corr}
\end{figure}
A representative set of correlation functions for $N=12$ neutrinos is shown in Fig.~\ref{fig:corr}, 
and a more complete set can be found in the SM.
For the two-body correlations, 
results from {\tt H1-1} and {\tt H1-1E}
follow the expected values, although somewhat limited by statistics.
This gets more prominent when looking at the three-body correlations, where with the current uncertainties most of the points are consistent with zero.
As seen by comparing the results from
{\tt H1-1} and {\tt H1-1E}, 
increasing the accumulated statistics by factor of 5 makes a substantial difference, 
but it remains insufficient.
Increasing further the number of shots would help resolve those small values, 
which would be reasonable with, for instance, IBM's quantum computers 
(where usually one works with $\gtrsim 10^4$ shots).
With the possibility of performing more shots, 
error mitigation techniques, 
such as randomized compiling~\cite{PhysRevA.94.052325} and decoherence renormalization~\cite{Urbanek:2021oej,Rahman:2022rlg,Farrell:2022wyt}, become viable.
While these correlations differ from zero, 
with a hierarchy $\mathcal{C}^{(2\nu)} > \mathcal{C}^{(3\nu)}$, 
an interesting trend is that $\mathcal{C}^{(2\nu)}_{1i}$ plateaus at late times, 
with $\mathcal{C}^{(2\nu)}_{12}$ reaching a value $\sim 0.5$, down to $\sim -0.5$ for $\mathcal{C}^{(2\nu)}_{1N}$. 
Other limiting  patterns are not found for other computed quantities.

A potential improvement to the current results has been explored, in which
a post-selection of the counts that satisfy the symmetries of the Hamiltonian. 
This technique has been used with great success when studying quantum field theories, 
e.g., see Refs.~\cite{Klco:2019evd,Nguyen:2021hyk,Farrell:2022wyt,Farrell:2022vyh}. 
However, specific to  this system in the flavor basis,
$\mathbf{b}\cdot \mathbf{J}$ is the conserved quantity (and not $J_z$), 
which makes post-selection not feasible in our studies.\footnote{Calculations could be performed in the mass basis, but the selected initial state would then be a sum of product states, 
which would complicate the calculation of the  $n$-tangle.}

\begin{figure}[!tb]
	\includegraphics[width=\columnwidth]{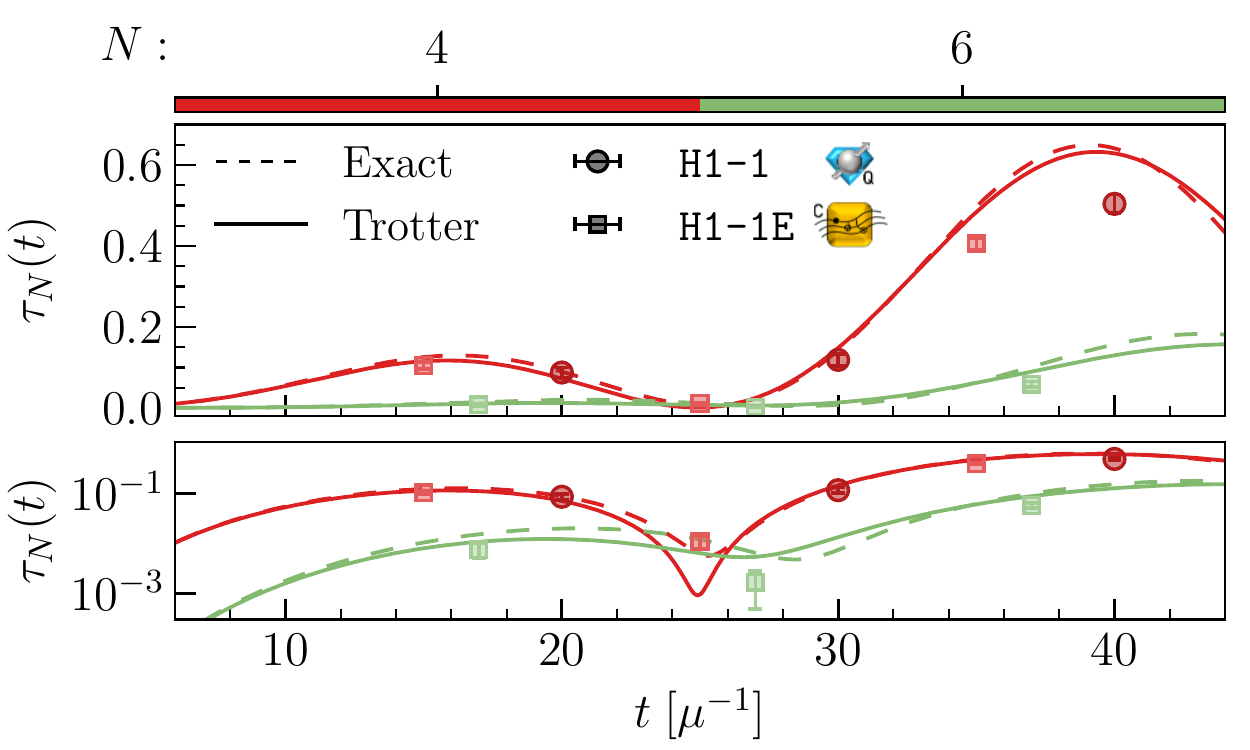}
	\caption{$N$-tangle $\tau_N$ for $N=\{4,6\}$ neutrinos on a linear (top) and log (bottom) scale. The lines show the Trotter (continuous) or exact (dashed) simulation, and the points show the results from {\tt H1-1} (dark circles, using 480 shots) and {\tt H1-1E} (light squares, using 1200 shots).}
	\label{fig:ntang1}
\end{figure}

\textit{Multi-Neutrino Entanglement ---} 
The $N$-tangle, $\tau_N$, is an interesting 
measure of multi-neutrino entanglement in these systems which
is straightforward to compute via,
\begin{equation}
    \tau_N (t)=|\langle \Psi_t|\sigma^{\otimes N}_y|\Psi_t^*\rangle |^2 =|\langle \Psi_0|e^{itH}\sigma^{\otimes N}_ye^{itH}|\Psi_0\rangle |^2 \ ,
\end{equation}
(without the need to use the SWAP test~\cite{Barenco:1996kz,Buhrman:2001,Beckey:2021mtb} to compute the overlap).
We have performed such calculations for the system of $N=4$ neutrinos 
using {\tt H1-1} and {\tt H1-1E}, as shown in
the top panel of Fig.~\ref{fig:ntang1}, 
where  two Trotter steps have been used in applying $e^{itH}$. 
The $N=6$ system has also been studied, but  only using {\tt H1-1E}
(and with a single Trotter step). 
For the time-range displayed, $\tau_N$ is seen to decrease rapidly with system size.

To further explore the dependence of multi-neutrino entanglement
with system size, 
it is interesting to look at $\tau_n$, with $n<N$. 
For this quantity, 
a re-scaling of the sum of the $\tau_n$s is found to be helpful,
\begin{equation}
    \tilde{\tau}_{n,N}=\frac{1}{N^{n-2}}\sum_i \tau^{(i)}_{n}\ ,
\label{eq:scaling}
\end{equation}
where the index $i$ in $\tau^{(i)}_{n}$ 
identifies one of the possible $\binom{N}{n}$ 
permutations of $\sigma^{\otimes n}_y$.\footnote{ 
e.g., for $N=4$ and $n=2$, these are $\sigma_y\otimes \sigma_y \otimes I \otimes I$, $\sigma_y\otimes I\otimes \sigma_y  \otimes I$,...}
\begin{figure}[!tb]
	\includegraphics[width=\columnwidth]{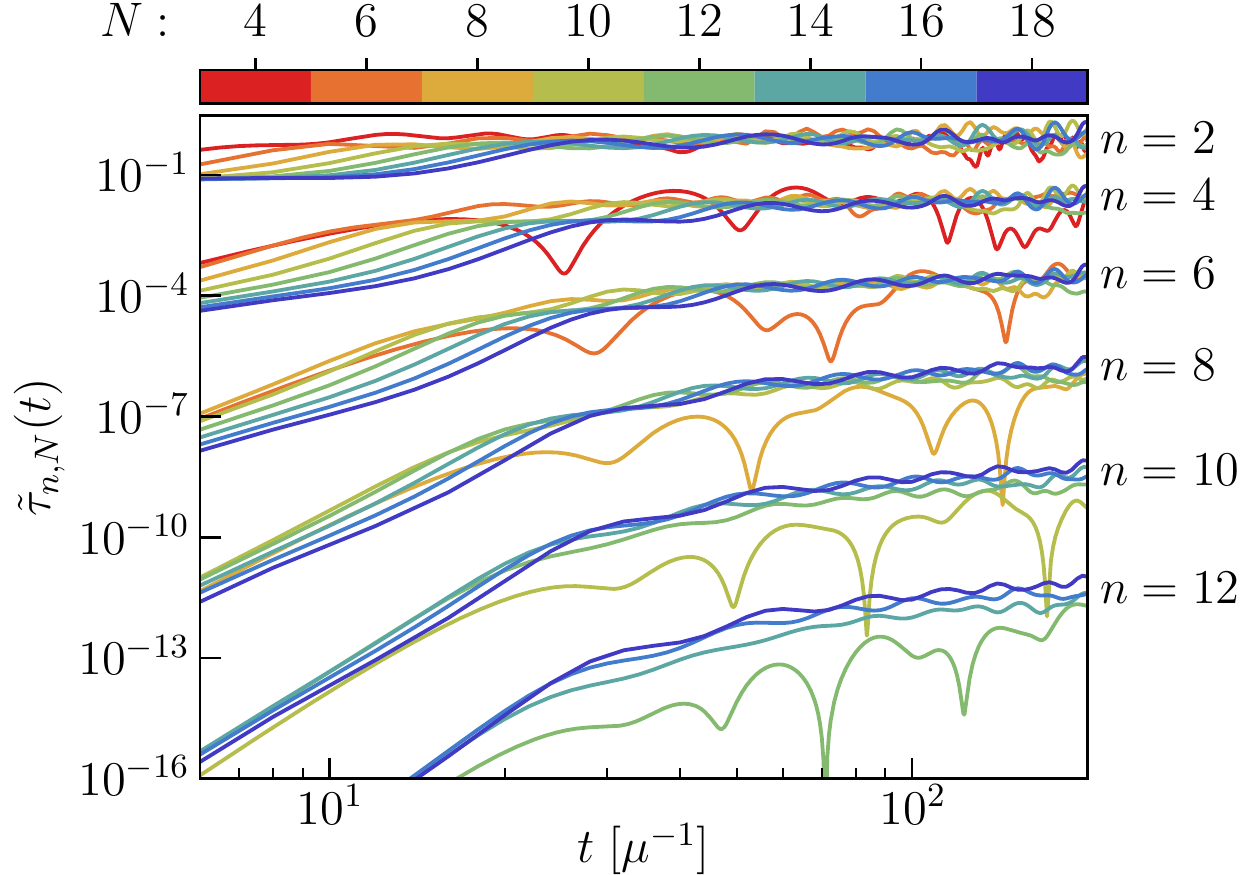}
	\caption{The re-scaled $n$-tangle, 
	$\tilde{\tau}_{n,N}$, defined in Eq.~(\ref{eq:scaling}), 
	for $n=\{2,\ldots,12\}$ and $N=\{4,\ldots,18\}$ neutrinos.}
	\label{fig:ntang2}
\end{figure}
Figure~\ref{fig:ntang2} shows the re-scaled $\tilde{\tau}_{n,N}$ 
for different values of $N$ and $n$.
These quantities are seen to exhibit convergence to a fixed curve with increasing $N$, starting for relatively small system sizes.
The time required to reach the plateau region increases with $n$, 
consistent with the notion that more time is required to  
entangle $n$ neutrinos than $n-1$.
Moreover, upon further investigation, 
these plateaus are found to be robust against variations of the two-body 
coupling strength (increasing or decreasing the angle of the cone) 
and vacuum-mixing angle (more or less oscillatory behavior), 
the only observed difference is in the relaxation time 
(shorter times for wider cones because the interaction strength $J_{ij}$ is stronger).

\begin{figure}[!tb]
	\includegraphics[width=\columnwidth]{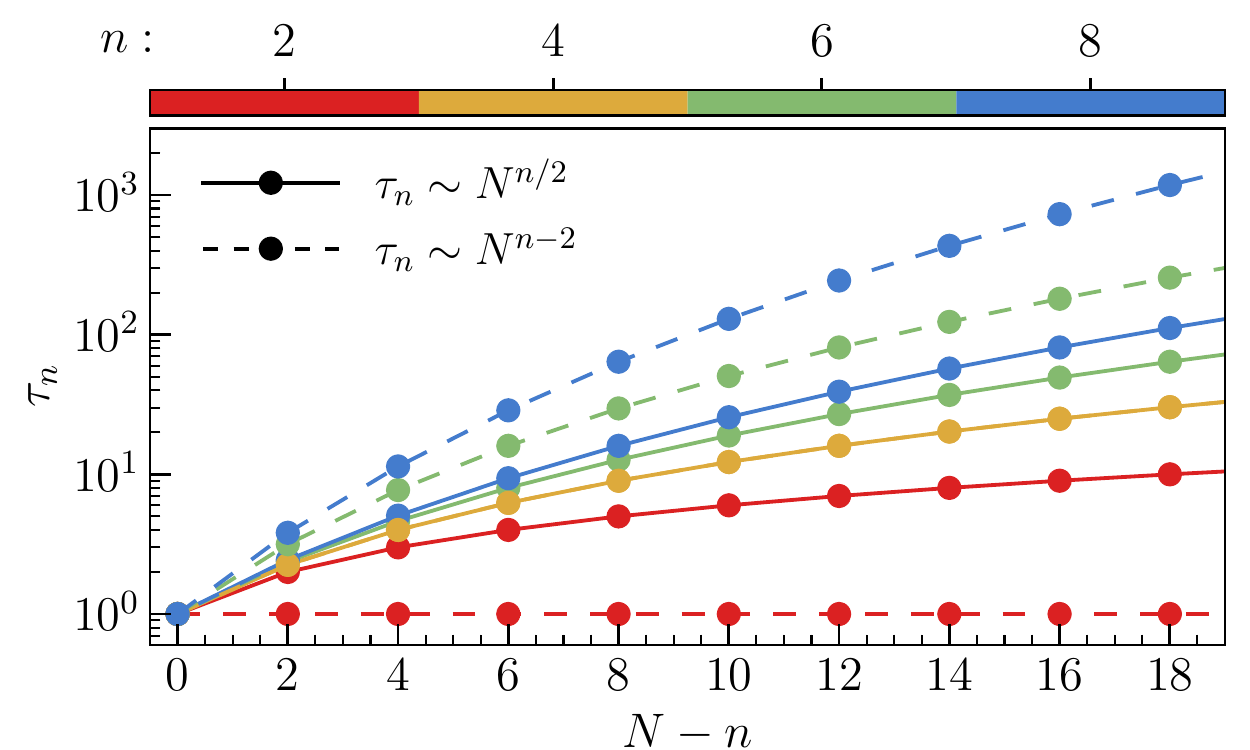}
	\caption{Comparison of two scalings of the $n$-tangles, 
	$N^{n/2}$ (continuous line) as found for systems of Bell pairs,
	and 
	$N^{n-2}$ (dashed line) as we have identified from the time evolution,
	for different values of $n$ and system sizes $N$, 
	normalized to $\tau_{n=N}=1$.}
	\label{fig:neutvsBellntan}
\end{figure}

While this scaling is somewhat  puzzling, 
it can be compared to 
other entangled multiqubit systems, for which analytic results are available. 
For example, the GHZ state has $\tau_{n=N}=1$ and $\tau_{n<N}=0$, while the W state has $\tau_{n}=0\; \forall n$ except for $\tau_{n=2}=2(N-1)/N$ (with $\tau_n=\sum_i \tau^{(i)}_n$ for $n<N$). 
A system that has similar scaling is the product of $N/2$ Bell pairs, with $\tau_{n}=\binom{N/2}{N/2-n/2}\sim N^{n/2}$. 
Comparing this with the scaling from Eq.~\eqref{eq:scaling}, 
the $n$-tangle grows faster for the neutrino state than the 
Bell-pairs-state with system size 
(except for $\tau_2$ and $\tau_4$), 
as depicted in Fig.~\ref{fig:neutvsBellntan}.
This implies  that there is multi-neutrino entanglement, 
beyond the two-neutrino entanglement found in the $(N/2)$-product Bell-pairs-state.

The present analysis concerns the coherent flavor oscillations of a mono-energetic neutrino gas
starting in a mixed-flavor pure state.
To gain an understanding of such dynamics in a mixed state, we have also examined the impact of different initial states for the evolution, and find that the values of $n$-tangle depend upon the initial state in non-trivial ways.
A mono-flavor initial state gives vanishing $n$-tangle, while an initial state with a single distinct flavor has non-zero $2$-tangle, resembling a W-state (the oscillations are too large to discriminate between the different $N$-scalings, consistent with $2(N-1)/N$ or $N^{n-2}$).
Gases of different energy neutrinos can be described within this framework through different values of ${\bf b}$ in Eq.~\eqref{eq:bdef}.

\textit{Summary and Conclusions ---} 
The time evolution of multi-neutrino quantum correlations and entanglement 
in dense neutrino systems is studied for systems of $N=4$, 6, 8 and 12 neutrinos using Quantinuum's {\tt H1-1} 20 qubit trapped ion quantum computer and its associated noisy emulator {\tt H1-1E}.
The central reason for including and developing 
methods for implementation on quantum computers is that determining the entanglement of the larger systems that will be required as input into realistic astrophysical simulations will need quantum simulations of such systems with larger numbers of neutrinos (than in this work). 
We have chosen to study the $n$-tangle in order to provide insight into the 
multi-particle entanglement structure of the systems. 
Compared to other entanglement witnesses, the $n$-tangle is straightforward to 
compute using a quantum computer (at least for a single product state, as considered here), 
since it does not require state tomography.
The behavior of re-scaled sums over $n$-tangles are found to converge to universal curves with increasing system sizes, that depend upon the parameters of the Hamiltonian,
with late-time plateaus.
Further, the magnitudes of the $n$-tangles 
are found to increasingly exceed that of systems comprised of neutrino Bell-pairs, 
indicating the presence of genuine multi-neutrino entanglement 
in collective coherent neutrino flavor oscillations in dense systems.

\vspace{1em}
\textit{Acknowledgements ---} We would like to thank Valentina Amitrano, Ramya Bhaskar, Joe Carlson, Anthony Ciavarella, Roland Farrell,
Francesco Pederiva, Alessandro Roggero and Francesco Turro, along with  
the QSC Thrust-2 team, for insightful discussions.  
This work was supported in part by the U.S.\ Department of Energy,
Office of Science, Office of Nuclear Physics, InQubator for Quantum Simulation (IQuS) 
(\url{https://iqus.uw.edu})
under Award Number DOE (NP) Award DE-SC0020970 (Savage), 
and the 
Quantum Science Center (QSC)
(\url{https://qscience.org}), 
a National Quantum Information Science Research Center of the U.S.\ Department of Energy (DOE) (Illa).
This work is also supported, in part, through the Department of Physics 
(\url{https://phys.washington.edu}) 
and the College of Arts and Sciences 
(\url{https://www.artsci.washington.edu})
at the University of Washington.
This research used resources of the Oak Ridge Leadership Computing Facility, 
which is a DOE Office of Science User Facility supported under Contract DE-AC05-00OR22725.
We have made extensive use of Wolfram {\tt Mathematica}~\cite{Mathematica},
{\tt Python}~\cite{python3,Hunter:2007},
{\tt Jupyter} Notebooks~\cite{PER-GRA:2007} 
in the {\tt Conda} environment~\cite{anaconda}, {\tt Julia}~\cite{Julia-2017,expokit,10.1145/285861.285868}
and the quantum programming environments: IBM's {\tt qiskit}~\cite{qiskit}
and CQC's {\tt pytket}~\cite{sivarajah2020t}.

\bibliography{bib}

\onecolumngrid
\clearpage
\newpage
\input{SM.tex}

\end{document}

%% file: SM.tex
\beginsupplement

\section{Supplementary Material\\ Multi-Neutrino Entanglement and Correlations in Dense  Neutrino Systems}

\noindent
This supplementary material contains additional figures and the tabulated results 
obtained with quantum simulations using Quantinuum's 20 trapped-ion quantum computer {\tt H1-1}, its noisy emulator {\tt H1-1E}, and other classical simulations,
used to produce the figures.
 In the legend of the figures, ``Exact'' results are obtained by computing the exponentiation of the Hamiltonian with {\tt Mathematica}~\cite{Mathematica}, except for the results for the $n$-tangle in Fig.~\ref{fig:ntang2}, where {\tt Julia}~\cite{Julia-2017} was used (with its {\tt Expokit} package~\cite{expokit}). The ``Trotter'' results were obtained with {\tt Mathematica} also, by first identifying the different terms in the Hamiltonian that can be implemented together (mainly, those shown in Eqs.~\eqref{eq:1bodyop} and~\eqref{eq:2bodyop}), 
 exponentiating them separately, and multiplying them in the order used in the quantum circuits 
 (as a check, this was compared with the results obtained with the {\tt qiskit} noiseless simulator~\cite{qiskit} using the quantum circuits run on {\tt H1-1}).
\begin{figure}[!htb]
	\includegraphics[width=0.85\columnwidth]{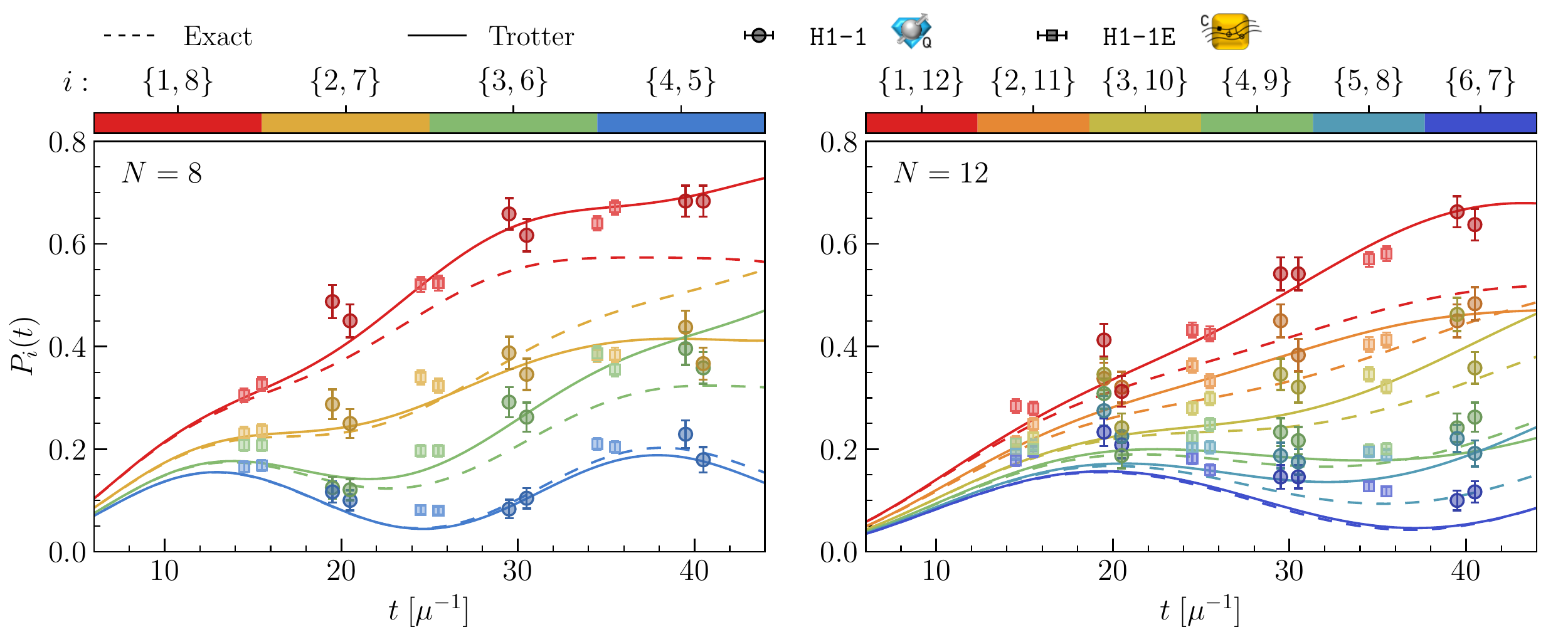}
	\caption{Flavor inversion probabilities for systems  with $N=8$ (left) and $N=12$ (right) neutrinos. 
	The lines show the single-step Trotter (continuous) and exact (dashed) simulations, 
	and the points show the results from {\tt H1-1} (dark circles, using 240 shots) 
	and {\tt H1-1E} (light squares, using 1200 shots).}
	\label{fig:time_ev_2}
\end{figure}
\begin{figure}[!htb]
	\includegraphics[width=0.85\columnwidth]{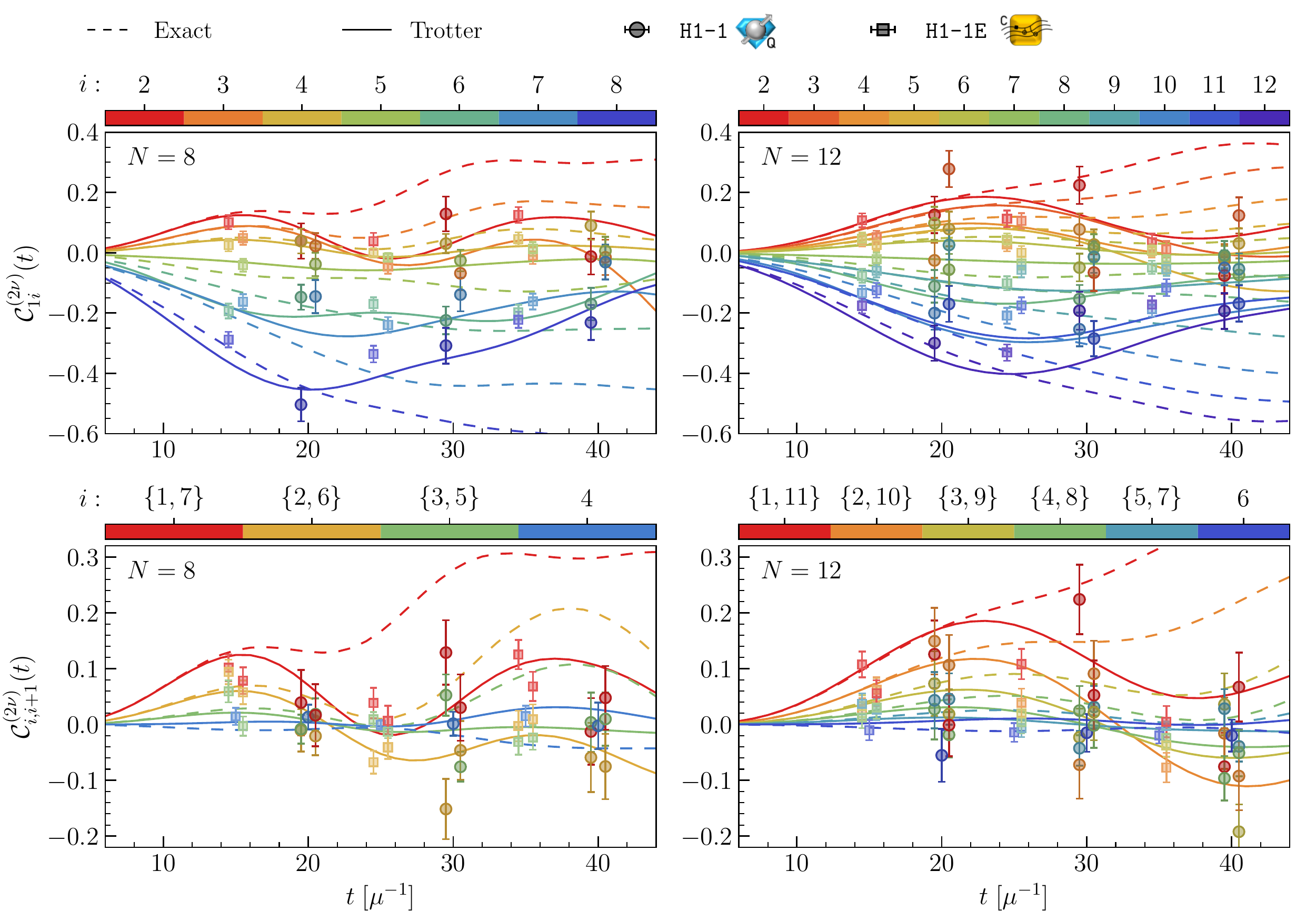}
	\caption{Two-body correlations for systems  with $N=8$ (left) and $N=12$ (right) neutrinos. 
	The lines show the single-step Trotter (continuous) and exact (dashed) simulations, 
	and the points show the results from {\tt H1-1} (dark circles, using 240 shots) 
	and {\tt H1-1E} (light squares, using 1200 shots).}
	\label{fig:corr_2}
\end{figure}
\begin{figure}[!htb]
	\includegraphics[width=0.85\columnwidth]{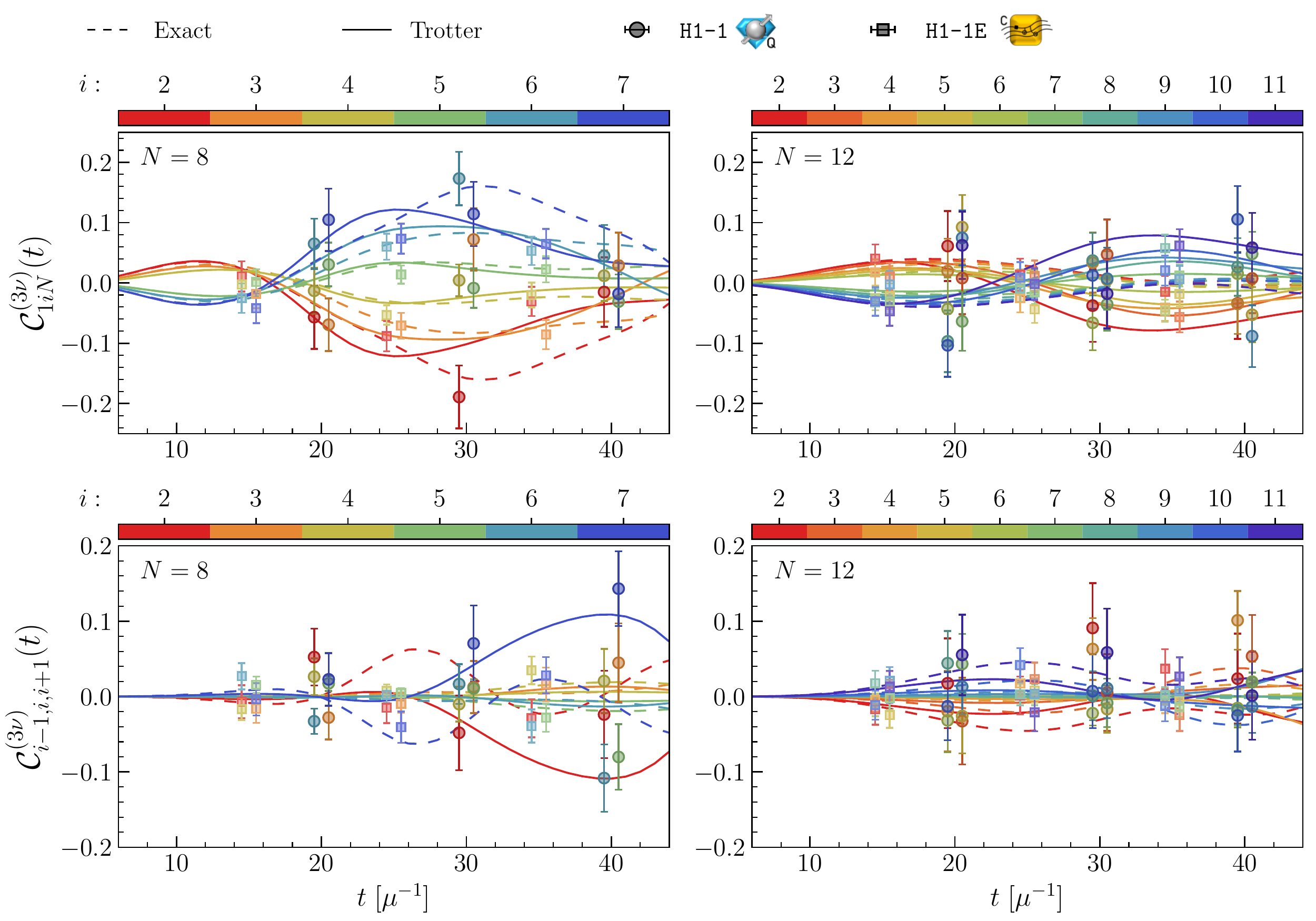}
	\caption{Three-body correlations for systems  with $N=8$ (left) and $N=12$ (right) neutrinos. 
	The lines show the single-step Trotter (continuous) and exact (dashed) simulations, 
	and the points show the results from {\tt H1-1} (dark circles, using 240 shots) 
	and {\tt H1-1E} (light squares, using 1200 shots).}
	\label{fig:corr_3}
\end{figure}
%
\begin{table}[!htb]
\centering
\begin{minipage}{0.48\columnwidth}
\centering
\begin{tabularx}{\columnwidth}{|| c | Y | Y | Y | Y | Y ||} 
\hline
 & \multicolumn{2}{c|}{$P_1(t)$} 
 & \multicolumn{2}{c|}{$P_8(t)$}
 & \\
 \hline
 $t$ & {\tt H1-1} & {\tt H1-1E} & {\tt H1-1} & {\tt H1-1E} & 1-T\\
 \hline\hline
 $15$ & $-$ & $0.305(13)$ & $-$ & $0.327(14)$ & $0.308$\\ 
 $20$ & $0.488(32)$ & $-$ & $0.450(32)$ & $-$ & $0.399$\\ 
 $25$ & $-$ & $0.521(14)$ & $-$ & $0.523(14)$ & $0.532$\\ 
 $30$ & $0.658(30)$ & $-$ & $0.617(31)$ & $-$ & $0.637$\\ 
 $35$ & $-$ & $0.640(14)$ & $-$ & $0.671(14)$ & $0.671$\\ 
 $40$ & $0.683(30)$ & $-$ & $0.683(30)$ & $-$ & $0.694$\\ 
 \hline
\end{tabularx}
\end{minipage}
\begin{minipage}{0.48\columnwidth}
\centering
\begin{tabularx}{\columnwidth}{|| c | Y | Y | Y | Y | Y ||} 
\hline
 & \multicolumn{2}{c|}{$P_2(t)$} 
 & \multicolumn{2}{c|}{$P_7(t)$}
 & \\
 \hline
 $t$ & {\tt H1-1} & {\tt H1-1E} & {\tt H1-1} & {\tt H1-1E} & 1-T\\
 \hline\hline
 $15$ & $-$ & $0.231(12)$ & $-$ & $0.236(12)$ & $0.226$\\ 
 $20$ & $0.287(29)$ & $-$ & $0.250(28)$ & $-$ & $0.242$\\ 
 $25$ & $-$ & $0.340(14)$ & $-$ & $0.323(14)$ & $0.291$\\ 
 $30$ & $0.387(31)$ & $-$ & $0.346(31)$ & $-$ & $0.361$\\ 
 $35$ & $-$ & $0.383(14)$ & $-$ & $0.383(14)$ & $0.406$\\ 
 $40$ & $0.438(32)$ & $-$ & $0.367(31)$ & $-$ & $0.415$\\ 
 \hline
\end{tabularx}
\end{minipage}
\vspace{1em}

\begin{minipage}{0.48\columnwidth}
\centering
\begin{tabularx}{\columnwidth}{|| c | Y | Y | Y | Y | Y ||} 
\hline
 & \multicolumn{2}{c|}{$P_3(t)$} 
 & \multicolumn{2}{c|}{$P_6(t)$}
 & \\
 \hline
 $t$ & {\tt H1-1} & {\tt H1-1E} & {\tt H1-1} & {\tt H1-1E} & 1-T\\
 \hline\hline
 $15$ & $-$ & $0.208(12)$ & $-$ & $0.208(12)$ & $0.175$\\ 
 $20$ & $0.125(21)$ & $-$ & $0.121(21)$ & $-$ & $0.145$\\ 
 $25$ & $-$ & $0.197(11)$ & $-$ & $0.197(12)$ & $0.163$\\ 
 $30$ & $0.292(29)$ & $-$ & $0.262(28)$ & $-$ & $0.258$\\ 
 $35$ & $-$ & $0.387(14)$ & $-$ & $0.355(14)$ & $0.360$\\ 
 $40$ & $0.396(32)$ & $-$ & $0.358(31)$ & $-$ & $0.424$\\ 
 \hline
\end{tabularx}
\end{minipage}
\begin{minipage}{0.48\columnwidth}
\centering
\begin{tabularx}{\columnwidth}{|| c | Y | Y | Y | Y | Y ||} 
\hline
 & \multicolumn{2}{c|}{$P_4(t)$} 
 & \multicolumn{2}{c|}{$P_5(t)$}
 & \\
 \hline
 $t$ & {\tt H1-1} & {\tt H1-1E} & {\tt H1-1} & {\tt H1-1E} & 1-T\\
 \hline\hline
 $15$ & $-$ & $0.165(11)$ & $-$ & $0.168(11)$ & $0.146$\\ 
 $20$ & $0.117(21)$ & $-$ & $0.100(19)$ & $-$ & $0.081$\\ 
 $25$ & $-$ & $0.082(8)$ & $-$ & $0.080(8)$ & $0.045$\\ 
 $30$ & $0.083(18)$ & $-$ & $0.104(20)$ & $-$ & $0.095$\\ 
 $35$ & $-$ & $0.210(12)$ & $-$ & $0.204(12)$ & $0.172$\\ 
 $40$ & $0.229(27)$ & $-$ & $0.179(25)$ & $-$ & $0.180$\\ 
 \hline
\end{tabularx}
\end{minipage}
\caption{
    Results from Fig.~\ref{fig:time_ev_2} showing the inversion probability $P_i(t)$ for the $N=8$ neutrino system. The results computed using {\tt H1-1} were obtained using 240 shots, and those  using {\tt H1-1E} with 1200 shots. The column labeled ``1-T'' is the exact value using a single Trotter step.
}
\end{table}
\begin{table}[!htb]
\centering
\begin{minipage}{0.48\columnwidth}
\centering
\begin{tabularx}{\columnwidth}{|| c | Y | Y | Y | Y | Y ||} 
\hline
 & \multicolumn{2}{c|}{$P_1(t)$} 
 & \multicolumn{2}{c|}{$P_{12}(t)$}
 & \\
 \hline
 $t$ & {\tt H1-1} & {\tt H1-1E} & {\tt H1-1} & {\tt H1-1E} & 1-T\\
 \hline\hline
 $15$ & $-$ & $0.284(13)$ & $-$ & $0.279(13)$ & $0.248$\\ 
 $20$ & $0.413(32)$ & $-$ & $0.313(30)$ & $-$ & $0.336$\\ 
 $25$ & $-$ & $0.432(14)$ & $-$ & $0.425(14)$ & $0.416$\\ 
 $30$ & $0.542(32)$ & $-$ & $0.542(32)$ & $-$ & $0.508$\\ 
 $35$ & $-$ & $0.570(14)$ & $-$ & $0.581(14)$ & $0.605$\\ 
 $40$ & $0.663(30)$ & $-$ & $0.638(31)$ & $-$ & $0.668$\\ 
 \hline
\end{tabularx}
\end{minipage}
\begin{minipage}{0.48\columnwidth}
\centering
\begin{tabularx}{\columnwidth}{|| c | Y | Y | Y | Y | Y ||} 
\hline
 & \multicolumn{2}{c|}{$P_2(t)$} 
 & \multicolumn{2}{c|}{$P_{11}(t)$}
 & \\
 \hline
 $t$ & {\tt H1-1} & {\tt H1-1E} & {\tt H1-1} & {\tt H1-1E} & 1-T\\
 \hline\hline
 $15$ & $-$ & $0.213(12)$ & $-$ & $0.248(12)$ & $0.212$\\ 
 $20$ & $0.338(31)$ & $-$ & $0.321(30)$ & $-$ & $0.282$\\ 
 $25$ & $-$ & $0.362(14)$ & $-$ & $0.333(14)$ & $0.333$\\ 
 $30$ & $0.450(32)$ & $-$ & $0.383(31)$ & $-$ & $0.385$\\ 
 $35$ & $-$ & $0.404(14)$ & $-$ & $0.413(14)$ & $0.433$\\ 
 $40$ & $0.450(32)$ & $-$ & $0.483(32)$ & $-$ & $0.462$\\ 
 \hline
\end{tabularx}
\end{minipage}
\vspace{1em}

\begin{minipage}{0.48\columnwidth}
\centering
\begin{tabularx}{\columnwidth}{|| c | Y | Y | Y | Y | Y ||} 
\hline
 & \multicolumn{2}{c|}{$P_3(t)$} 
 & \multicolumn{2}{c|}{$P_{10}(t)$}
 & \\
 \hline
 $t$ & {\tt H1-1} & {\tt H1-1E} & {\tt H1-1} & {\tt H1-1E} & 1-T\\
 \hline\hline
 $15$ & $-$ & $0.198(11)$ & $-$ & $0.223(12)$ & $0.178$\\ 
 $20$ & $0.346(31)$ & $-$ & $0.242(28)$ & $-$ & $0.224$\\ 
 $25$ & $-$ & $0.280(13)$ & $-$ & $0.299(13)$ & $0.245$\\ 
 $30$ & $0.346(31)$ & $-$ & $0.321(30)$ & $-$ & $0.269$\\ 
 $35$ & $-$ & $0.346(14)$ & $-$ & $0.322(14)$ & $0.321$\\ 
 $40$ & $0.463(32)$ & $-$ & $0.358(31)$ & $-$ & $0.399$\\ 
 \hline
\end{tabularx}
\end{minipage}
\begin{minipage}{0.48\columnwidth}
\centering
\begin{tabularx}{\columnwidth}{|| c | Y | Y | Y | Y | Y ||} 
\hline
 & \multicolumn{2}{c|}{$P_4(t)$} 
 & \multicolumn{2}{c|}{$P_9(t)$}
 & \\
 \hline
 $t$ & {\tt H1-1} & {\tt H1-1E} & {\tt H1-1} & {\tt H1-1E} & 1-T\\
 \hline\hline
 $15$ & $-$ & $0.211(12)$ & $-$ & $0.200(12)$ & $0.161$\\ 
 $20$ & $0.308(30)$ & $-$ & $0.188(25)$ & $-$ & $0.196$\\ 
 $25$ & $-$ & $0.223(12)$ & $-$ & $0.248(12)$ & $0.197$\\ 
 $30$ & $0.233(27)$ & $-$ & $0.217(27)$ & $-$ & $0.184$\\ 
 $35$ & $-$ & $0.198(11)$ & $-$ & $0.200(12)$ & $0.178$\\ 
 $40$ & $0.242(28)$ & $-$ & $0.262(28)$ & $-$ & $0.193$\\ 
 \hline
\end{tabularx}
\end{minipage}
\vspace{1em}

\begin{minipage}{0.48\columnwidth}
\centering
\begin{tabularx}{\columnwidth}{|| c | Y | Y | Y | Y | Y ||} 
\hline
 & \multicolumn{2}{c|}{$P_5(t)$} 
 & \multicolumn{2}{c|}{$P_8(t)$}
 & \\
 \hline
 $t$ & {\tt H1-1} & {\tt H1-1E} & {\tt H1-1} & {\tt H1-1E} & 1-T\\
 \hline\hline
 $15$ & $-$ & $0.191(11)$ & $-$ & $0.208(12)$ & $0.146$\\ 
 $20$ & $0.275(29)$ & $-$ & $0.225(27)$ & $-$ & $0.172$\\ 
 $25$ & $-$ & $0.203(12)$ & $-$ & $0.203(12)$ & $0.160$\\ 
 $30$ & $0.187(25)$ & $-$ & $0.175(25)$ & $-$ & $0.139$\\ 
 $35$ & $-$ & $0.195(11)$ & $-$ & $0.189(11)$ & $0.142$\\ 
 $40$ & $0.221(27)$ & $-$ & $0.192(25)$ & $-$ & $0.186$\\ 
 \hline
\end{tabularx}
\end{minipage}
\begin{minipage}{0.48\columnwidth}
\centering
\begin{tabularx}{\columnwidth}{|| c | Y | Y | Y | Y | Y ||} 
\hline
 & \multicolumn{2}{c|}{$P_6(t)$} 
 & \multicolumn{2}{c|}{$P_7(t)$}
 & \\
 \hline
 $t$ & {\tt H1-1} & {\tt H1-1E} & {\tt H1-1} & {\tt H1-1E} & 1-T\\
 \hline\hline
 $15$ & $-$ & $0.178(11)$ & $-$ & $0.196(11)$ & $0.138$\\ 
 $20$ & $0.233(27)$ & $-$ & $0.208(26)$ & $-$ & $0.157$\\ 
 $25$ & $-$ & $0.182(11)$ & $-$ & $0.159(11)$ & $0.132$\\ 
 $30$ & $0.146(23)$ & $-$ & $0.146(23)$ & $-$ & $0.085$\\ 
 $35$ & $-$ & $0.127(10)$ & $-$ & $0.118(9)$ & $0.050$\\ 
 $40$ & $0.100(19)$ & $-$ & $0.117(21)$ & $-$ & $0.053$\\ 
 \hline
\end{tabularx}
\end{minipage}
\caption{
    Results from Fig.~\ref{fig:time_ev_2} showing the inversion probability $P_i(t)$ for the $N=12$ neutrino system. The results computed using {\tt H1-1} were obtained using 240 shots, and those using {\tt H1-1E} with 1200 shots. The column labeled ``1-T'' is the exact value using a single Trotter step.
}
\end{table}
\begin{table}[!htb]
\centering
\begin{minipage}{0.32\columnwidth}
\centering
\begin{tabularx}{\columnwidth}{|| c | Y | Y | Y ||} 
\hline
 & \multicolumn{3}{c||}{$\mathcal{C}^{(2\nu)}_{1,2}(t)$} \\
 \hline
 $t$ & {\tt H1-1} & {\tt H1-1E} & 1-T\\
 \hline\hline
 $15$ & $-$ & $0.102(24)$ & $0.125$ \\ 
 $20$ & $0.039(58)$ & $-$ & $0.071$ \\ 
 $25$ & $-$ & $0.038(27)$ & $-0.017$\\ 
 $30$ & $0.129(58)$ & $-$ & $0.033$ \\ 
 $35$ & $-$ & $0.125(26)$ & $0.111$ \\ 
 $40$ & $-0.012(60)$ & $-$ & $0.105$\\ 
 \hline
\end{tabularx}
\end{minipage}
\begin{minipage}{0.32\columnwidth}
\centering
\begin{tabularx}{\columnwidth}{|| c | Y | Y | Y ||} 
\hline
 & \multicolumn{3}{c||}{$\mathcal{C}^{(2\nu)}_{1,3}(t)$} \\
 \hline
 $t$ & {\tt H1-1} & {\tt H1-1E} & 1-T\\
 \hline\hline
 $15$ & $-$ & $0.049(22)$ & $0.088$\\ 
 $20$ & $0.023(43)$ & $-$ & $0.050$\\ 
 $25$ & $-$ & $-0.046(23)$ & $-0.033$\\ 
 $30$ & $-0.068(57)$ & $-$ & $-0.016$\\ 
 $35$ & $-$ & $-0.010(27)$ & $0.043$\\ 
 $40$ & $-0.015(59)$ & $-$ & $-0.022$\\ 
 \hline
\end{tabularx}
\end{minipage}
\begin{minipage}{0.32\columnwidth}
\centering
\begin{tabularx}{\columnwidth}{|| c | Y | Y | Y ||} 
\hline
 & \multicolumn{3}{c||}{$\mathcal{C}^{(2\nu)}_{1,4}(t)$} \\
 \hline
 $t$ & {\tt H1-1} & {\tt H1-1E} & 1-T\\
 \hline\hline
 $15$ & $-$ & $0.025(20)$ & $0.043$\\ 
 $20$ & $0.039(41)$ & $-$ & $0.019$\\ 
 $25$ & $-$ & $0.000(16)$ & $-0.013$\\ 
 $30$ & $0.030(32)$ & $-$ & $0.003$\\ 
 $35$ & $-$ & $0.046(22)$ & $0.023$\\ 
 $40$ & $0.090(47)$ & $-$ & $0.021$\\ 
 \hline
\end{tabularx}
\end{minipage}
\vspace{1em}

\begin{minipage}{0.32\columnwidth}
\centering
\begin{tabularx}{\columnwidth}{|| c | Y | Y | Y ||} 
\hline
 & \multicolumn{3}{c||}{$\mathcal{C}^{(2\nu)}_{1,5}(t)$} \\
 \hline
 $t$ & {\tt H1-1} & {\tt H1-1E} & 1-T\\
 \hline\hline
 $15$ & $-$ & $-0.041(21)$ & $-0.030$\\ 
 $20$ & $-0.038(39)$ & $-$ & $-0.050$\\ 
 $25$ & $-$ & $-0.017(16)$ & $-0.058$\\ 
 $30$ & $-0.025(36)$ & $-$ & $-0.045$\\ 
 $35$ & $-$ & $0.013(22)$ & $-0.029$\\ 
 $40$ & $0.006(46)$ & $-$ & $-0.021$\\ 
 \hline
\end{tabularx}
\end{minipage}
\begin{minipage}{0.32\columnwidth}
\centering
\begin{tabularx}{\columnwidth}{|| c | Y | Y | Y ||} 
\hline
 & \multicolumn{3}{c||}{$\mathcal{C}^{(2\nu)}_{1,6}(t)$} \\
 \hline
 $t$ & {\tt H1-1} & {\tt H1-1E} & 1-T\\
 \hline\hline
 $15$ & $-$ & $-0.193(24)$ & $-0.187$\\ 
 $20$ & $-0.147(41)$ & $-$ & $-0.211$\\ 
 $25$ & $-$ & $-0.170(22)$ & $-0.199$\\ 
 $30$ & $-0.224(46)$ & $-$ & $-0.220$\\ 
 $35$ & $-$ & $-0.198(25)$ & $-0.216$\\ 
 $40$ & $-0.170(54)$ & $-$ & $-0.147$\\ 
 \hline
\end{tabularx}
\end{minipage}
\begin{minipage}{0.32\columnwidth}
\centering
\begin{tabularx}{\columnwidth}{|| c | Y | Y | Y ||} 
\hline
 & \multicolumn{3}{c||}{$\mathcal{C}^{(2\nu)}_{1,7}(t)$} \\
 \hline
 $t$ & {\tt H1-1} & {\tt H1-1E} & 1-T\\
 \hline\hline
 $15$ & $-$ & $-0.162(24)$ & $-0.188$\\ 
 $20$ & $-0.145(55)$ & $-$ & $-0.264$\\ 
 $25$ & $-$ & $-0.239(26)$ & $-0.268$\\ 
 $30$ & $-0.139(55)$ & $-$ & $-0.216$\\ 
 $35$ & $-$ & $-0.162(26)$ & $-0.165$\\ 
 $40$ & $-0.031(57)$ & $-$ & $-0.130$\\ 
 \hline
\end{tabularx}
\end{minipage}
\vspace{1em}

\begin{minipage}{0.32\columnwidth}
\centering
\begin{tabularx}{\columnwidth}{|| c | Y | Y | Y ||} 
\hline
 & \multicolumn{3}{c||}{$\mathcal{C}^{(2\nu)}_{1,8}(t)$} \\
 \hline
 $t$ & {\tt H1-1} & {\tt H1-1E} & 1-T\\
 \hline\hline
 $15$ & $-$ & $-0.288(26)$ & $-0.378$\\ 
 $20$ & $-0.504(56)$ & $-$ & $-0.454$\\ 
 $25$ & $-$ & $-0.336(27)$ & $-0.409$\\ 
 $30$ & $-0.308(59)$ & $-$ & $-0.346$\\ 
 $35$ & $-$ & $-0.222(27)$ & $-0.279$\\ 
 $40$ & $-0.231(58)$ & $-$ & $-0.174$\\ 
 \hline
\end{tabularx}
\end{minipage}
\caption{
    Results from Fig.~\ref{fig:corr_2} showing the two-body correlation $\mathcal{C}^{(2\nu)}_{1i}(t)$ for the $N=8$ neutrino system. The results computed using {\tt H1-1} were obtained using 240 shots, and those using {\tt H1-1E} with 1200 shots. The column labeled ``1-T'' is the exact value using a single Trotter step.
}
\end{table}
\begin{table}[!htb]
\centering
\begin{minipage}{0.48\columnwidth}
\centering
\begin{tabularx}{\columnwidth}{|| c | Y | Y | Y | Y | c ||} 
\hline
 & \multicolumn{2}{c|}{$\mathcal{C}^{(2\nu)}_{1,2}(t)$} 
 & \multicolumn{2}{c|}{$\mathcal{C}^{(2\nu)}_{7,8}(t)$}
 & \\
 \hline
 $t$ & {\tt H1-1} & {\tt H1-1E} & {\tt H1-1} & {\tt H1-1E} & 1-T\\
 \hline\hline
 $15$ & $-$ & $0.102(24)$ & $-$ & $0.079(24)$ & $0.125$ \\ 
 $20$ & $0.039(58)$ & $-$ & $0.017(56)$ & $-$ & $0.071$ \\ 
 $25$ & $-$ & $0.038(27)$ & $-$ & $0.007(27)$ & $-0.017$\\ 
 $30$ & $0.129(58)$ & $-$ & $0.030(59)$ & $-$ & $0.033$ \\ 
 $35$ & $-$ & $0.125(26)$ & $-$ & $0.068(26)$ & $0.111$ \\ 
 $40$ & $-0.012(60)$ & $-$& $0.048(57)$ & $-$ &  $0.105$\\ 
 \hline
\end{tabularx}
\end{minipage}
\begin{minipage}{0.48\columnwidth}
\centering
\begin{tabularx}{\columnwidth}{|| c | Y | Y | Y | Y | c ||} 
\hline
 & \multicolumn{2}{c|}{$\mathcal{C}^{(2\nu)}_{2,3}(t)$} 
 & \multicolumn{2}{c|}{$\mathcal{C}^{(2\nu)}_{6,7}(t)$}
 & \\
 \hline
 $t$ & {\tt H1-1} & {\tt H1-1E} & {\tt H1-1} & {\tt H1-1E} & 1-T\\
 \hline\hline
 $15$ & $-$ & $0.094(22)$ & $-$ & $0.058(21)$ & $0.060$\\ 
 $20$ & $-0.011(38)$ & $-$ & $-0.021(34)$ & $-$ & $0.023$\\ 
 $25$ & $-$ & $-0.067(21)$ & $-$ & $-0.041(21)$ & $-0.053$\\ 
 $30$ & $-0.151(54)$ & $-$ & $-0.046(53)$ & $-$ & $-0.053$\\ 
 $35$ & $-$ & $-0.003(27)$ & $-$ & $0.009(27)$ & $-0.020$\\ 
 $40$ & $-0.059(62)$ & $-$ & $-0.075(58)$ & $-$ & $-0.044$\\ 
 \hline
\end{tabularx}
\end{minipage}
\vspace{1em}

\begin{minipage}{0.48\columnwidth}
\centering
\begin{tabularx}{\columnwidth}{|| c | Y | Y | Y | Y | c ||} 
\hline
 & \multicolumn{2}{c|}{$\mathcal{C}^{(2\nu)}_{3,4}(t)$} 
 & \multicolumn{2}{c|}{$\mathcal{C}^{(2\nu)}_{5,6}(t)$}
 & \\
 \hline
 $t$ & {\tt H1-1} & {\tt H1-1E} & {\tt H1-1} & {\tt H1-1E} & 1-T\\
 \hline\hline
 $15$ & $-$ & $0.059(19)$ & $-$ & $-0.003(17)$ & $0.021$\\ 
 $20$ & $-0.008(26)$ & $-$ & $0.018(29)$ & $-$ & $0.008$\\ 
 $25$ & $-$ & $0.009(13)$ & $-$ & $-0.013(11)$ & $-0.012$\\ 
 $30$ & $0.053(37)$ & $-$ & $-0.076(27)$ & $-$ & $-0.010$\\ 
 $35$ & $-$ & $-0.031(23)$ & $-$ & $-0.023(22)$ & $-0.005$\\ 
 $40$ & $0.004(53)$ & $-$ & $0.010(48)$ & $-$ & $-0.010$\\ 
 \hline
\end{tabularx}
\end{minipage}
\begin{minipage}{0.34\columnwidth}
\centering
\begin{tabularx}{\columnwidth}{|| c | Y | Y | Y ||} 
\hline
 & \multicolumn{3}{c||}{$\mathcal{C}^{(2\nu)}_{4,5}(t)$} \\
 \hline
 $t$ & {\tt H1-1} & {\tt H1-1E} & 1-T\\
 \hline\hline
 $15$ & $-$ & $0.015(15)$ & $0.004$\\ 
 $20$ & $0.013(22)$ & $-$ & $0.004$\\ 
 $25$ & $-$ & $-0.001(9)$ & $-0.003$\\ 
 $30$ & $0.001(21)$ & $-$ & $0.007$\\ 
 $35$ & $-$ & $0.015(19)$ & $0.028$\\ 
 $40$ & $-0.002(42)$ & $-$ & $0.026$\\ 
 \hline
\end{tabularx}
\end{minipage}
\caption{
    Results from Fig.~\ref{fig:corr_2} showing the two-body correlation $\mathcal{C}^{(2\nu)}_{i,i+1}(t)$ for the $N=8$ neutrino system. The results computed using {\tt H1-1} were obtained using 240 shots, and those using {\tt H1-1E} with 1200 shots. The column labeled ``1-T'' is the exact value using a single Trotter step.
}
\end{table}
\begin{table}[!htb]
\centering
\begin{minipage}{0.32\columnwidth}
\centering
\begin{tabularx}{\columnwidth}{|| c | Y | Y | Y ||} 
\hline
 & \multicolumn{3}{c||}{$\mathcal{C}^{(3\nu)}_{1,2,8}(t)$} \\
 \hline
 $t$ & {\tt H1-1} & {\tt H1-1E} & 1-T\\
 \hline\hline
 $15$ & $-$ & $0.011(25)$ & $0.019$ \\ 
 $20$ & $-0.057(53)$ & $-$ & $-0.068$ \\ 
 $25$ & $-$ & $-0.088(25)$ & $-0.122$\\ 
 $30$ & $-0.189(52)$ & $-$ & $-0.102$ \\ 
 $35$ & $-$ & $-0.031(24)$ & $-0.062$ \\ 
 $40$ & $-0.015(58)$ & $-$ & $-0.034$\\ 
 \hline
\end{tabularx}
\end{minipage}
\begin{minipage}{0.32\columnwidth}
\centering
\begin{tabularx}{\columnwidth}{|| c | Y | Y | Y ||} 
\hline
 & \multicolumn{3}{c||}{$\mathcal{C}^{(3\nu)}_{1,3,8}(t)$} \\
 \hline
 $t$ & {\tt H1-1} & {\tt H1-1E} & 1-T\\
 \hline\hline
 $15$ & $-$ & $-0.018(24)$ & $0.018$\\ 
 $20$ & $-0.069(44)$ & $-$ & $-0.040$\\ 
 $25$ & $-$ & $-0.071(21)$ & $-0.089$\\ 
 $30$ & $0.072(52)$ & $-$ & $-0.093$\\ 
 $35$ & $-$ & $-0.086(24)$ & $-0.072$\\ 
 $40$ & $0.029(54)$ & $-$ & $-0.025$\\ 
 \hline
\end{tabularx}
\end{minipage}
\begin{minipage}{0.32\columnwidth}
\centering
\begin{tabularx}{\columnwidth}{|| c | Y | Y | Y ||} 
\hline
 & \multicolumn{3}{c||}{$\mathcal{C}^{(3\nu)}_{1,4,8}(t)$} \\
 \hline
 $t$ & {\tt H1-1} & {\tt H1-1E} & 1-T\\
 \hline\hline
 $15$ & $-$ & $-0.003(20)$ & $0.020$\\ 
 $20$ & $-0.013(38)$ & $-$ & $-0.010$\\ 
 $25$ & $-$ & $-0.053(16)$ & $-0.034$\\ 
 $30$ & $0.004(26)$ & $-$ & $-0.024$\\ 
 $35$ & $-$ & $-0.020(20)$ & $-0.012$\\ 
 $40$ & $0.012(42)$ & $-$ & $-0.008$\\ 
 \hline
\end{tabularx}
\end{minipage}
\vspace{1em}

\begin{minipage}{0.32\columnwidth}
\centering
\begin{tabularx}{\columnwidth}{|| c | Y | Y | Y ||} 
\hline
 & \multicolumn{3}{c||}{$\mathcal{C}^{(3\nu)}_{1,5,8}(t)$} \\
 \hline
 $t$ & {\tt H1-1} & {\tt H1-1E} & 1-T\\
 \hline\hline
 $15$ & $-$ & $0.001(21)$ & $-0.020$\\ 
 $20$ & $0.031(36)$ & $-$ & $0.010$\\ 
 $25$ & $-$ & $0.014(15)$ & $0.034$\\ 
 $30$ & $-0.009(33)$ & $-$ & $0.024$\\ 
 $35$ & $-$ & $0.023(22)$ & $0.012$\\ 
 $40$ & $-0.031(45)$ & $-$ & $0.008$\\ 
 \hline
\end{tabularx}
\end{minipage}
\begin{minipage}{0.32\columnwidth}
\centering
\begin{tabularx}{\columnwidth}{|| c | Y | Y | Y ||} 
\hline
 & \multicolumn{3}{c||}{$\mathcal{C}^{(3\nu)}_{1,6,8}(t)$} \\
 \hline
 $t$ & {\tt H1-1} & {\tt H1-1E} & 1-T\\
 \hline\hline
 $15$ & $-$ & $-0.025(25)$ & $-0.018$\\ 
 $20$ & $0.065(42)$ & $-$ & $0.040$\\ 
 $25$ & $-$ & $0.060(22)$ & $0.089$\\ 
 $30$ & $0.173(44)$ & $-$ & $0.093$\\ 
 $35$ & $-$ & $0.053(23)$ & $0.072$\\ 
 $40$ & $0.045(51)$ & $-$ & $0.025$\\ 
 \hline
\end{tabularx}
\end{minipage}
\begin{minipage}{0.32\columnwidth}
\centering
\begin{tabularx}{\columnwidth}{|| c | Y | Y | Y ||} 
\hline
 & \multicolumn{3}{c||}{$\mathcal{C}^{(3\nu)}_{1,7,8}(t)$} \\
 \hline
 $t$ & {\tt H1-1} & {\tt H1-1E} & 1-T\\
 \hline\hline
 $15$ & $-$ & $-0.042(25)$ & $-0.019$\\ 
 $20$ & $0.105(52)$ & $-$ & $0.068$\\ 
 $25$ & $-$ & $0.073(25)$ & $0.122$\\ 
 $30$ & $0.114(54)$ & $-$ & $0.102$\\ 
 $35$ & $-$ & $0.065(24)$ & $0.062$\\ 
 $40$ & $-0.018(55)$ & $-$ & $0.034$\\ 
 \hline
\end{tabularx}
\end{minipage}
\caption{
    Results from Fig.~\ref{fig:corr_3} showing the three-body correlation $\mathcal{C}^{(3\nu)}_{1iN}(t)$ for the $N=8$ neutrino system. The results computed using {\tt H1-1} were obtained using 240 shots, and those using {\tt H1-1E} with 1200 shots. The column labeled ``1-T'' is the exact value using a single Trotter step.
}
\end{table}
\begin{table}[!htb]
\centering
\begin{minipage}{0.32\columnwidth}
\centering
\begin{tabularx}{\columnwidth}{|| c | Y | Y | Y ||} 
\hline
 & \multicolumn{3}{c||}{$\mathcal{C}^{(3\nu)}_{1,2,3}(t)$} \\
 \hline
 $t$ & {\tt H1-1} & {\tt H1-1E} & 1-T\\
 \hline\hline
 $15$ & $-$ & $-0.007(22)$ & $-0.003$\\ 
 $20$ & $0.052(38)$ & $-$ & $0.001$\\ 
 $25$ & $-$ & $-0.015(20)$ & $0.004$\\ 
 $30$ & $-0.048(50)$ & $-$ & $-0.036$\\ 
 $35$ & $-$ & $-0.029(25)$ & $-0.088$\\ 
 $40$ & $-0.024(58)$ & $-$ & $-0.109$\\ 
 \hline
\end{tabularx}
\end{minipage}
\begin{minipage}{0.32\columnwidth}
\centering
\begin{tabularx}{\columnwidth}{|| c | Y | Y | Y ||} 
\hline
 & \multicolumn{3}{c||}{$\mathcal{C}^{(3\nu)}_{2,3,4}(t)$} \\
 \hline
 $t$ & {\tt H1-1} & {\tt H1-1E} & 1-T\\
 \hline\hline
 $15$ & $-$ & $-0.016(19)$ & $0.001$\\ 
 $20$ & $-0.028(29)$ & $-$ & $0.001$\\ 
 $25$ & $-$ & $-0.009(12)$ & $-0.001$\\ 
 $30$ & $0.013(35)$ & $-$ & $0.001$\\ 
 $35$ & $-$ & $0.016(22)$ & $0.009$\\ 
 $40$ & $0.045(52)$ & $-$ & $0.013$\\ 
 \hline
\end{tabularx}
\end{minipage}
\begin{minipage}{0.32\columnwidth}
\centering
\begin{tabularx}{\columnwidth}{|| c | Y | Y | Y ||} 
\hline
 & \multicolumn{3}{c||}{$\mathcal{C}^{(3\nu)}_{3,4,5}(t)$} \\
 \hline
 $t$ & {\tt H1-1} & {\tt H1-1E} & 1-T\\
 \hline\hline
 $15$ & $-$ & $-0.017(14)$ & $0.0005$\\ 
 $20$ & $0.026(24)$ & $-$ & $0.001$\\ 
 $25$ & $-$ & $0.007(9)$ & $0.0001$\\ 
 $30$ & $-0.010(23)$ & $-$ & $0.002$\\ 
 $35$ & $-$ & $0.035(18)$ & $0.005$\\ 
 $40$ & $0.021(43)$ & $-$ & $0.006$\\ 
 \hline
\end{tabularx}
\end{minipage}
\vspace{1em}

\begin{minipage}{0.32\columnwidth}
\centering
\begin{tabularx}{\columnwidth}{|| c | Y | Y | Y ||} 
\hline
 & \multicolumn{3}{c||}{$\mathcal{C}^{(3\nu)}_{4,5,6}(t)$} \\
 \hline
 $t$ & {\tt H1-1} & {\tt H1-1E} & 1-T\\
 \hline\hline
 $15$ & $-$ & $0.016(11)$ & $-0.0005$\\ 
 $20$ & $0.018(10)$ & $-$ & $-0.001$\\ 
 $25$ & $-$ & $0.005(6)$ & $-0.0001$\\ 
 $30$ & $0.011(12)$ & $-$ & $-0.002$\\ 
 $35$ & $-$ & $-0.028(19)$ & $-0.005$\\ 
 $40$ & $-0.080(43)$ & $-$ & $-0.006$\\ 
 \hline
\end{tabularx}
\end{minipage}
\begin{minipage}{0.32\columnwidth}
\centering
\begin{tabularx}{\columnwidth}{|| c | Y | Y | Y ||} 
\hline
 & \multicolumn{3}{c||}{$\mathcal{C}^{(3\nu)}_{5,6,7}(t)$} \\
 \hline
 $t$ & {\tt H1-1} & {\tt H1-1E} & 1-T\\
 \hline\hline
 $15$ & $-$ & $0.027(17)$ & $-0.001$\\ 
 $20$ & $-0.033(17)$ & $-$ & $-0.001$\\ 
 $25$ & $-$ & $0.002(10)$ & $0.001$\\ 
 $30$ & $0.017(26)$ & $-$ & $-0.001$\\ 
 $35$ & $-$ & $-0.039(22)$ & $-0.009$\\ 
 $40$ & $-0.108(45)$ & $-$ & $-0.013$\\ 
 \hline
\end{tabularx}
\end{minipage}
\begin{minipage}{0.32\columnwidth}
\centering
\begin{tabularx}{\columnwidth}{|| c | Y | Y | Y ||} 
\hline
 & \multicolumn{3}{c||}{$\mathcal{C}^{(3\nu)}_{6,7,8}(t)$} \\
 \hline
 $t$ & {\tt H1-1} & {\tt H1-1E} & 1-T\\
 \hline\hline
 $15$ & $-$ & $-0.003(21)$ & $0.003$ \\  
 $20$ & $0.023(35)$ & $-$ & $-0.001$ \\ 
 $25$ & $-$ & $-0.041(21)$ & $-0.004$\\ 
 $30$ & $0.070(50)$ & $-$ & $0.036$ \\ 
 $35$ & $-$ & $0.028(24)$ & $0.088$ \\ 
 $40$ & $0.143(49)$ & $-$ & $0.109$\\ 
 \hline
\end{tabularx}
\end{minipage}
\caption{
    Results from Fig.~\ref{fig:corr_3} showing the three-body correlation $\mathcal{C}^{(3\nu)}_{i-1,i,i+1}(t)$ for the $N=8$ neutrino system. The results computed using {\tt H1-1} were obtained using 240 shots, and those using {\tt H1-1E} with 1200 shots. The column labeled ``1-T'' is the exact value using a single Trotter step.
}
\end{table}
\begin{table}[!htb]
\centering
\begin{minipage}{0.32\columnwidth}
\centering
\begin{tabularx}{\columnwidth}{|| c | Y | Y | Y ||} 
\hline
 & \multicolumn{3}{c||}{$\mathcal{C}^{(2\nu)}_{1,2}(t)$} \\
 \hline
 $t$ & {\tt H1-1} & {\tt H1-1E} & 1-T\\
 \hline\hline
 $15$ & $-$ & $0.107(23)$ & $0.107$ \\ 
 $20$ & $0.126(61)$ & $-$ & $0.173$ \\ 
 $25$ & $-$ & $0.113(28)$ & $0.179$\\ 
 $30$ & $0.224(62)$ & $-$ & $0.119$ \\ 
 $35$ & $-$ & $0.035(28)$ & $0.057$ \\ 
 $40$ & $-0.076(61)$ & $-$ & $0.054$\\ 
 \hline
\end{tabularx}
\end{minipage}
\begin{minipage}{0.32\columnwidth}
\centering
\begin{tabularx}{\columnwidth}{|| c | Y | Y | Y ||} 
\hline
 & \multicolumn{3}{c||}{$\mathcal{C}^{(2\nu)}_{1,3}(t)$} \\
 \hline
 $t$ & {\tt H1-1} & {\tt H1-1E} & 1-T\\
 \hline\hline
 $15$ & $-$ & $0.051(22)$ & $0.084$\\ 
 $20$ & $0.278(60)$ & $-$ & $0.141$\\ 
 $25$ & $-$ & $0.105(26)$ & $0.155$\\ 
 $30$ & $-0.066(61)$ & $-$ & $0.111$\\ 
 $35$ & $-$ & $0.012(27)$ & $0.040$\\ 
 $40$ & $0.124(60)$ & $-$ & $-0.007$\\ 
 \hline
\end{tabularx}
\end{minipage}
\begin{minipage}{0.32\columnwidth}
\centering
\begin{tabularx}{\columnwidth}{|| c | Y | Y | Y ||} 
\hline
 & \multicolumn{3}{c||}{$\mathcal{C}^{(2\nu)}_{1,4}(t)$} \\
 \hline
 $t$ & {\tt H1-1} & {\tt H1-1E} & 1-T\\
 \hline\hline
 $15$ & $-$ & $0.044(22)$ & $0.060$\\ 
 $20$ & $-0.025(58)$ & $-$ & $0.094$\\ 
 $25$ & $-$ & $0.028(24)$ & $0.088$\\ 
 $30$ & $0.077(53)$ & $-$ & $0.043$\\ 
 $35$ & $-$ & $0.006(23)$ & $-0.001$\\ 
 $40$ & $-0.024(53)$ & $-$ & $-0.003$\\ 
 \hline
\end{tabularx}
\end{minipage}
\vspace{1em}

\begin{minipage}{0.32\columnwidth}
\centering
\begin{tabularx}{\columnwidth}{|| c | Y | Y | Y ||} 
\hline
 & \multicolumn{3}{c||}{$\mathcal{C}^{(2\nu)}_{1,5}(t)$} \\
 \hline
 $t$ & {\tt H1-1} & {\tt H1-1E} & 1-T\\
 \hline\hline
 $15$ & $-$ & $0.026(21)$ & $0.045$\\ 
 $20$ & $0.079(57)$ & $-$ & $0.076$\\ 
 $25$ & $-$ & $0.000(23)$ & $0.077$\\ 
 $30$ & $0.027(50)$ & $-$ & $0.031$\\ 
 $35$ & $-$ & $-0.051(23)$ & $-0.048$\\ 
 $40$ & $0.031(50)$ & $-$ & $-0.112$\\ 
 \hline
\end{tabularx}
\end{minipage}
\begin{minipage}{0.32\columnwidth}
\centering
\begin{tabularx}{\columnwidth}{|| c | Y | Y | Y ||} 
\hline
 & \multicolumn{3}{c||}{$\mathcal{C}^{(2\nu)}_{1,6}(t)$} \\
 \hline
 $t$ & {\tt H1-1} & {\tt H1-1E} & 1-T\\
 \hline\hline
 $15$ & $-$ & $0.047(21)$ & $0.027$\\ 
 $20$ & $0.098(55)$ & $-$ & $0.039$\\ 
 $25$ & $-$ & $0.042(22)$ & $0.034$\\ 
 $30$ & $-0.049(46)$ & $-$ & $0.013$\\ 
 $35$ & $-$ & $0.009(19)$ & $-0.004$\\ 
 $40$ & $-0.015(37)$ & $-$ & $-0.005$\\ 
 \hline
\end{tabularx}
\end{minipage}
\begin{minipage}{0.32\columnwidth}
\centering
\begin{tabularx}{\columnwidth}{|| c | Y | Y | Y ||} 
\hline
 & \multicolumn{3}{c||}{$\mathcal{C}^{(2\nu)}_{1,7}(t)$} \\
 \hline
 $t$ & {\tt H1-1} & {\tt H1-1E} & 1-T\\
 \hline\hline
 $15$ & $-$ & $-0.021(21)$ & $-0.014$\\ 
 $20$ & $-0.056(52)$ & $-$ & $-0.022$\\ 
 $25$ & $-$ & $-0.035(21)$ & $-0.030$\\ 
 $30$ & $0.016(45)$ & $-$ & $-0.036$\\ 
 $35$ & $-$ & $-0.025(18)$ & $-0.036$\\ 
 $40$ & $-0.074(35)$ & $-$ & $-0.033$\\ 
 \hline
\end{tabularx}
\end{minipage}
\vspace{1em}

\begin{minipage}{0.32\columnwidth}
\centering
\begin{tabularx}{\columnwidth}{|| c | Y | Y | Y ||} 
\hline
 & \multicolumn{3}{c||}{$\mathcal{C}^{(2\nu)}_{1,8}(t)$} \\
 \hline
 $t$ & {\tt H1-1} & {\tt H1-1E} & 1-T\\
 \hline\hline
 $15$ & $-$ & $-0.076(23)$ & $-0.095$\\ 
 $20$ & $-0.111(54)$ & $-$ & $-0.151$\\ 
 $25$ & $-$ & $-0.101(23)$ & $-0.169$\\ 
 $30$ & $-0.153(46)$ & $-$ & $-0.144$\\ 
 $35$ & $-$ & $-0.049(22)$ & $-0.108$\\ 
 $40$ & $-0.009(48)$ & $-$ & $-0.084$\\ 
 \hline
\end{tabularx}
\end{minipage}
\begin{minipage}{0.32\columnwidth}
\centering
\begin{tabularx}{\columnwidth}{|| c | Y | Y | Y ||} 
\hline
 & \multicolumn{3}{c||}{$\mathcal{C}^{(2\nu)}_{1,9}(t)$} \\
 \hline
 $t$ & {\tt H1-1} & {\tt H1-1E} & 1-T\\
 \hline\hline
 $15$ & $-$ & $-0.059(22)$ & $-0.061$\\ 
 $20$ & $0.026(49)$ & $-$ & $-0.094$\\ 
 $25$ & $-$ & $-0.057(25)$ & $-0.119$\\ 
 $30$ & $-0.014(53)$ & $-$ & $-0.126$\\ 
 $35$ & $-$ & $-0.054(22)$ & $-0.114$\\ 
 $40$ & $-0.054(52)$ & $-$ & $-0.098$\\ 
 \hline
\end{tabularx}
\end{minipage}
\begin{minipage}{0.32\columnwidth}
\centering
\begin{tabularx}{\columnwidth}{|| c | Y | Y | Y ||} 
\hline
 & \multicolumn{3}{c||}{$\mathcal{C}^{(2\nu)}_{1,10}(t)$} \\
 \hline
 $t$ & {\tt H1-1} & {\tt H1-1E} & 1-T\\
 \hline\hline
 $15$ & $-$ & $-0.134(24)$ & $-0.162$\\ 
 $20$ & $-0.200(55)$ & $-$ & $-0.253$\\ 
 $25$ & $-$ & $-0.209(26)$ & $-0.296$\\ 
 $30$ & $-0.254(56)$ & $-$ & $-0.281$\\ 
 $35$ & $-$ & $-0.186(26)$ & $-0.238$\\ 
 $40$ & $-0.050(58)$ & $-$ & $-0.197$\\ 
 \hline
\end{tabularx}
\end{minipage}
\vspace{1em}

\begin{minipage}{0.32\columnwidth}
\centering
\begin{tabularx}{\columnwidth}{|| c | Y | Y | Y ||} 
\hline
 & \multicolumn{3}{c||}{$\mathcal{C}^{(2\nu)}_{1,11}(t)$} \\
 \hline
 $t$ & {\tt H1-1} & {\tt H1-1E} & 1-T\\
 \hline\hline
 $15$ & $-$ & $-0.124(24)$ & $-0.160$\\ 
 $20$ & $-0.170(59)$ & $-$ & $-0.241$\\ 
 $25$ & $-$ & $-0.174(27)$ & $-0.283$\\ 
 $30$ & $-0.285(59)$ & $-$ & $-0.264$\\ 
 $35$ & $-$ & $-0.116(28)$ & $-0.208$\\ 
 $40$ & $-0.169(60)$ & $-$ & $-0.166$\\ 
 \hline
\end{tabularx}
\end{minipage}
\begin{minipage}{0.32\columnwidth}
\centering
\begin{tabularx}{\columnwidth}{|| c | Y | Y | Y ||} 
\hline
 & \multicolumn{3}{c||}{$\mathcal{C}^{(2\nu)}_{1,12}(t)$} \\
 \hline
 $t$ & {\tt H1-1} & {\tt H1-1E} & 1-T\\
 \hline\hline
 $15$ & $-$ & $-0.176(25)$ & $-0.246$\\ 
 $20$ & $-0.300(58)$ & $-$ & $-0.360$\\ 
 $25$ & $-$ & $-0.331(27)$ & $-0.402$\\ 
 $30$ & $-0.192(63)$ & $-$ & $-0.359$\\ 
 $35$ & $-$ & $-0.172(28)$ & $-0.278$\\ 
 $40$ & $-0.193(60)$ & $-$ & $-0.218$\\ 
 \hline
\end{tabularx}
\end{minipage}
\caption{
    Results from Fig.~\ref{fig:corr_2} showing the two-body correlation $\mathcal{C}^{(2\nu)}_{1i}(t)$ for the $N=12$ neutrino system. The results computed using {\tt H1-1} were obtained using 240 shots, and those using {\tt H1-1E} with 1200 shots. The column labeled ``1-T'' is the exact value using a single Trotter step.
}
\end{table}
\begin{table}[!htb]
\centering
\begin{minipage}{0.48\columnwidth}
\centering
\begin{tabularx}{\columnwidth}{|| c | Y | Y | Y | Y | c ||} 
\hline
 & \multicolumn{2}{c|}{$\mathcal{C}^{(2\nu)}_{1,2}(t)$} 
 & \multicolumn{2}{c|}{$\mathcal{C}^{(2\nu)}_{11,12}(t)$}
 & \\
 \hline
 $t$ & {\tt H1-1} & {\tt H1-1E} & {\tt H1-1} & {\tt H1-1E} & 1-T\\
 \hline\hline
 $15$ & $-$ & $0.107(23)$ & $-$ & $0.056(23)$ & $0.107$ \\ 
 $20$ & $0.126(61)$ & $-$ & $-0.001(56)$ & $-$ & $0.173$ \\ 
 $25$ & $-$ & $0.113(28)$ & $-$ & $0.108(27)$ & $0.179$ \\ 
 $30$ & $0.224(62)$ & $-$ & $0.052(62)$ & $-$ & $0.119$ \\ 
 $35$ & $-$ & $0.035(28)$ & $-$ & $0.005(28)$ & $0.057$ \\ 
 $40$ & $-0.076(61)$ & $-$& $0.067(62)$ & $-$ &  $0.054$\\ 
 \hline
\end{tabularx}
\end{minipage}
\begin{minipage}{0.48\columnwidth}
\centering
\begin{tabularx}{\columnwidth}{|| c | Y | Y | Y | Y | c ||} 
\hline
 & \multicolumn{2}{c|}{$\mathcal{C}^{(2\nu)}_{2,3}(t)$} 
 & \multicolumn{2}{c|}{$\mathcal{C}^{(2\nu)}_{10,11}(t)$}
 & \\
 \hline
 $t$ & {\tt H1-1} & {\tt H1-1E} & {\tt H1-1} & {\tt H1-1E} & 1-T\\
 \hline\hline
 $15$ & $-$ & $0.027(20)$ & $-$ & $0.062(22)$ & $0.068$\\ 
 $20$ & $0.149(59)$ & $-$ & $0.106(54)$ & $-$ & $0.111$\\ 
 $25$ & $-$ & $0.051(25)$ & $-$ & $0.039(25)$ & $0.108$\\ 
 $30$ & $-0.072(61)$ & $-$ & $0.091(59)$ & $-$ & $0.041$\\ 
 $35$ & $-$ & $-0.082(27)$ & $-$ & $-0.077(26)$ & $-0.054$\\ 
 $40$ & $-0.016(64)$ & $-$ & $-0.092(61)$ & $-$ & $-0.109$\\ 
 \hline
\end{tabularx}
\end{minipage}
\vspace{1em}

\begin{minipage}{0.48\columnwidth}
\centering
\begin{tabularx}{\columnwidth}{|| c | Y | Y | Y | Y | c ||} 
\hline
 & \multicolumn{2}{c|}{$\mathcal{C}^{(2\nu)}_{3,4}(t)$} 
 & \multicolumn{2}{c|}{$\mathcal{C}^{(2\nu)}_{9,10}(t)$}
 & \\
 \hline
 $t$ & {\tt H1-1} & {\tt H1-1E} & {\tt H1-1} & {\tt H1-1E} & 1-T\\
 \hline\hline
 $15$ & $-$ & $0.033(20)$ & $-$ & $0.062(21)$ & $0.038$\\ 
 $20$ & $0.073(58)$ & $-$ & $0.019(44)$ & $-$ & $0.060$\\ 
 $25$ & $-$ & $0.024(22)$ & $-$ & $0.020(23)$ & $0.054$\\ 
 $30$ & $-0.023(51)$ & $-$ & $0.022(50)$ & $-$ & $0.012$\\ 
 $35$ & $-$ & $-0.023(21)$ & $-$ & $-0.037(21)$ & $-0.039$\\ 
 $40$ & $0.036(55)$ & $-$ & $-0.192(49)$ & $-$ & $-0.060$\\ 
 \hline
\end{tabularx}
\end{minipage}
\begin{minipage}{0.48\columnwidth}
\centering
\begin{tabularx}{\columnwidth}{|| c | Y | Y | Y | Y | c ||} 
\hline
 & \multicolumn{2}{c|}{$\mathcal{C}^{(2\nu)}_{4,5}(t)$} 
 & \multicolumn{2}{c|}{$\mathcal{C}^{(2\nu)}_{8,9}(t)$}
 & \\
 \hline
 $t$ & {\tt H1-1} & {\tt H1-1E} & {\tt H1-1} & {\tt H1-1E} & 1-T\\
 \hline\hline
 $15$ & $-$ & $0.012(19)$ & $-$ & $0.030(20)$ & $0.020$\\ 
 $20$ & $0.027(54)$ & $-$ & $-0.019(41)$ & $-$ & $0.031$\\ 
 $25$ & $-$ & $0.013(20)$ & $-$ & $0.005(20)$ & $0.024$\\ 
 $30$ & $0.025(44)$ & $-$ & $-0.002(40)$ & $-$ & $-0.0002$\\ 
 $35$ & $-$ & $-0.007(18)$ & $-$ & $-0.025(17)$ & $-0.027$\\ 
 $40$ & $-0.096(40)$ & $-$ & $-0.051(42)$ & $-$ & $-0.040$\\ 
 \hline
\end{tabularx}
\end{minipage}
\vspace{1em}

\begin{minipage}{0.48\columnwidth}
\centering
\begin{tabularx}{\columnwidth}{|| c | Y | Y | Y | Y | c ||} 
\hline
 & \multicolumn{2}{c|}{$\mathcal{C}^{(2\nu)}_{5,6}(t)$} 
 & \multicolumn{2}{c|}{$\mathcal{C}^{(2\nu)}_{7,8}(t)$}
 & \\
 \hline
 $t$ & {\tt H1-1} & {\tt H1-1E} & {\tt H1-1} & {\tt H1-1E} & 1-T\\
 \hline\hline
 $15$ & $-$ & $0.037(19)$ & $-$ & $0.027(19)$ & $0.009$\\ 
 $20$ & $0.043(50)$ & $-$ & $0.046(46)$ & $-$ & $0.013$\\ 
 $25$ & $-$ & $0.019(18)$ & $-$ & $-0.006(17)$ & $0.008$\\ 
 $30$ & $-0.043(31)$ & $-$ & $0.031(38)$ & $-$ & $-0.001$\\ 
 $35$ & $-$ & $0.011(16)$ & $-$ & $-0.002(14)$ & $-0.008$\\ 
 $40$ & $0.028(35)$ & $-$ & $-0.039(27)$ & $-$ & $-0.011$\\ 
 \hline
\end{tabularx}
\end{minipage}
\begin{minipage}{0.34\columnwidth}
\centering
\begin{tabularx}{\columnwidth}{|| c | Y | Y | Y ||} 
\hline
 & \multicolumn{3}{c||}{$\mathcal{C}^{(2\nu)}_{6,7}(t)$} \\
 \hline
 $t$ & {\tt H1-1} & {\tt H1-1E} & 1-T\\
 \hline\hline
 $15$ & $-$ & $-0.010(18)$ & $0.002$\\ 
 $20$ & $-0.055(47)$ & $-$ & $0.007$\\ 
 $25$ & $-$ & $-0.014(17)$ & $0.011$\\ 
 $30$ & $-0.015(34)$ & $-$ & $0.008$\\ 
 $35$ & $-$ & $-0.020(14)$ & $0.001$\\ 
 $40$ & $-0.020(29)$ & $-$ & $0.0005$\\ 
 \hline
\end{tabularx}
\end{minipage}
\caption{
    Results from Fig.~\ref{fig:corr_2} showing the two-body correlation $\mathcal{C}^{(2\nu)}_{i,i+1}(t)$ for the $N=12$ neutrino system. The results computed using {\tt H1-1} were obtained using 240 shots, and those using {\tt H1-1E} with 1200 shots. The column labeled ``1-T'' is the exact value using a single Trotter step.
}
\end{table}
\begin{table}[!htb]
\centering
\begin{minipage}{0.32\columnwidth}
\centering
\begin{tabularx}{\columnwidth}{|| c | Y | Y | Y ||} 
\hline
 & \multicolumn{3}{c||}{$\mathcal{C}^{(3\nu)}_{1,2,12}(t)$} \\
 \hline
 $t$ & {\tt H1-1} & {\tt H1-1E} & 1-T\\
 \hline\hline
 $15$ & $-$ & $0.040(24)$ &  $0.034$ \\ 
 $20$ & $0.061(58)$ & $-$ & $0.020$ \\ 
 $25$ & $-$ & $0.014(26)$ & $-0.025$\\ 
 $30$ & $-0.038(60)$ & $-$ & $-0.069$ \\ 
 $35$ & $-$ & $-0.015(27)$ & $-0.078$ \\ 
 $40$ & $-0.034(59)$ & $-$ & $-0.062$\\ 
 \hline
\end{tabularx}
\end{minipage}
\begin{minipage}{0.32\columnwidth}
\centering
\begin{tabularx}{\columnwidth}{|| c | Y | Y | Y ||} 
\hline
 & \multicolumn{3}{c||}{$\mathcal{C}^{(3\nu)}_{1,3,12}(t)$} \\
 \hline
 $t$ & {\tt H1-1} & {\tt H1-1E} & 1-T\\
 \hline\hline
 $15$ & $-$ & $0.010(23)$ & $0.032$ \\ 
 $20$ & $0.008(60)$ & $-$ & $0.031$ \\ 
 $25$ & $-$ & $0.012(25)$ & $0.003$ \\ 
 $30$ & $0.047(58)$ & $-$ &  $-0.036$ \\ 
 $35$ & $-$ & $-0.056(26)$ & $-0.054$\\ 
 $40$ & $0.007(58)$ & $-$ &  $-0.039$ \\ 
 \hline
\end{tabularx}
\end{minipage}
\begin{minipage}{0.32\columnwidth}
\centering
\begin{tabularx}{\columnwidth}{|| c | Y | Y | Y ||} 
\hline
 & \multicolumn{3}{c||}{$\mathcal{C}^{(3\nu)}_{1,4,12}(t)$} \\
 \hline
 $t$ & {\tt H1-1} & {\tt H1-1E} & 1-T\\
 \hline\hline
 $15$ & $-$ & $0.017(22)$ & $0.024$ \\ 
 $20$ & $0.019(54)$ & $-$ & $0.019$ \\ 
 $25$ & $-$ & $-0.025(23)$ & $-0.005$\\ 
 $30$ & $0.034(50)$ & $-$ & $-0.034$ \\ 
 $35$ & $-$ & $-0.041(22)$ & $-0.042$\\ 
 $40$ & $-0.035(50)$ & $-$ & $-0.033$ \\ 
 \hline
\end{tabularx}
\end{minipage}
\vspace{1em}

\begin{minipage}{0.32\columnwidth}
\centering
\begin{tabularx}{\columnwidth}{|| c | Y | Y | Y ||} 
\hline
 & \multicolumn{3}{c||}{$\mathcal{C}^{(3\nu)}_{1,5,12}(t)$} \\
 \hline
 $t$ & {\tt H1-1} & {\tt H1-1E} & 1-T\\
 \hline\hline
 $15$ & $-$ & $-0.022(21)$ & $0.020$  \\ 
 $20$ & $0.094(53)$ & $-$ & $0.020$  \\ 
 $25$ & $-$ & $-0.044(23)$ & $0.040$  \\ 
 $30$ & $0.005(47)$ & $-$ & $-0.022$\\ 
 $35$ & $-$ & $-0.018(21)$ & $-0.036$ \\ 
 $40$ & $-0.053(46)$ & $-$ & $-0.025$\\ 
 \hline
\end{tabularx}
\end{minipage}
\begin{minipage}{0.32\columnwidth}
\centering
\begin{tabularx}{\columnwidth}{|| c | Y | Y | Y ||} 
\hline
 & \multicolumn{3}{c||}{$\mathcal{C}^{(3\nu)}_{1,6,12}(t)$} \\
 \hline
 $t$ & {\tt H1-1} & {\tt H1-1E} & 1-T\\
 \hline\hline
 $15$ & $-$ & $-0.026(20)$ & $0.013$\\ 
 $20$ & $-0.043(53)$ & $-$ & $0.013$\\ 
 $25$ & $-$ & $0.001(21)$ & $0.003$\\ 
 $30$ & $-0.066(45)$ & $-$ & $-0.009$\\ 
 $35$ & $-$ & $-0.046(18)$ & $-0.015$\\ 
 $40$ & $0.015(36)$ & $-$ & $-0.013$\\ 
 \hline
\end{tabularx}
\end{minipage}
\begin{minipage}{0.32\columnwidth}
\centering
\begin{tabularx}{\columnwidth}{|| c | Y | Y | Y ||} 
\hline
 & \multicolumn{3}{c||}{$\mathcal{C}^{(3\nu)}_{1,7,12}(t)$} \\
 \hline
 $t$ & {\tt H1-1} & {\tt H1-1E} & 1-T\\
 \hline\hline
 $15$ & $-$ & $-0.031(21)$ & $-0.013$\\ 
 $20$ & $-0.064(49)$ & $-$ & $-0.013$\\ 
 $25$ & $-$ & $-0.015(20)$ & $-0.003$\\ 
 $30$ & $-0.035(44)$ & $-$ & $0.009$\\ 
 $35$ & $-$ & $0.012(18)$ & $0.015$\\ 
 $40$ & $0.048(36)$ & $-$ &$0.013$\\ 
 \hline
\end{tabularx}
\end{minipage}
\vspace{1em}

\begin{minipage}{0.32\columnwidth}
\centering
\begin{tabularx}{\columnwidth}{|| c | Y | Y | Y ||} 
\hline
 & \multicolumn{3}{c||}{$\mathcal{C}^{(3\nu)}_{1,8,12}(t)$} \\
 \hline
 $t$ & {\tt H1-1} & {\tt H1-1E} & 1-T\\
 \hline\hline
 $15$ & $-$ & $-0.032(22)$ & $-0.020$  \\ 
 $20$ & $-0.097(51)$ & $-$ & $-0.020$  \\ 
 $25$ & $-$ & $-0.009(22)$ & $-0.040$  \\ 
 $30$ & $0.037(46)$ & $-$ & $0.022$\\ 
 $35$ & $-$ & $0.058(22)$ & $0.036$ \\ 
 $40$ & $0.026(48)$ & $-$ & $0.025$\\ 
 \hline
\end{tabularx}
\end{minipage}
\begin{minipage}{0.32\columnwidth}
\centering
\begin{tabularx}{\columnwidth}{|| c | Y | Y | Y ||} 
\hline
 & \multicolumn{3}{c||}{$\mathcal{C}^{(3\nu)}_{1,9,12}(t)$} \\
 \hline
 $t$ & {\tt H1-1} & {\tt H1-1E} & 1-T\\
 \hline\hline
 $15$ & $-$ & $-0.002(22)$ & $-0.024$ \\ 
 $20$ & $0.075(45)$ & $-$ &  $-0.019$ \\ 
 $25$ & $-$ & $-0.001(24)$ &  $0.005$\\ 
 $30$ & $0.007(51)$ & $-$ & $0.034$ \\ 
 $35$ & $-$ & $0.007(22)$ &  $0.042$\\ 
 $40$ & $-0.088(51)$ & $-$ & $0.033$ \\ 
 \hline
\end{tabularx}
\end{minipage}
\begin{minipage}{0.32\columnwidth}
\centering
\begin{tabularx}{\columnwidth}{|| c | Y | Y | Y ||} 
\hline
 & \multicolumn{3}{c||}{$\mathcal{C}^{(3\nu)}_{1,10,12}(t)$} \\
 \hline
 $t$ & {\tt H1-1} & {\tt H1-1E} & 1-T\\
 \hline\hline
 $15$ & $-$ & $-0.031(24)$ &$-0.032$  \\ 
 $20$ & $-0.103(52)$ & $-$ & $-0.031$  \\ 
 $25$ & $-$ & $0.009(26)$ & $-0.003$ \\ 
 $30$ & $0.013(55)$ & $-$ & $0.036$\\ 
 $35$ & $-$ & $0.021(25)$ & $0.054$\\ 
 $40$ & $0.105(55)$ & $-$ &$0.039$\\ 
 \hline
\end{tabularx}
\end{minipage}
\vspace{1em}

\begin{minipage}{0.32\columnwidth}
\centering
\begin{tabularx}{\columnwidth}{|| c | Y | Y | Y ||} 
\hline
 & \multicolumn{3}{c||}{$\mathcal{C}^{(3\nu)}_{1,11,12}(t)$} \\
 \hline
 $t$ & {\tt H1-1} & {\tt H1-1E} & 1-T\\
 \hline\hline
 $15$ & $-$ & $-0.047(24)$ & $-0.034$\\ 
 $20$ & $0.062(56)$ & $-$ & $-0.020$\\
 $25$ & $-$ & $-0.002(26)$ & $0.025$ \\
 $30$ & $-0.018(58)$ & $-$ &$0.069$ \\ 
 $35$ & $-$ & $0.062(27)$ & $0.078$ \\
 $40$ & $0.059(58)$ & $-$ &$0.062$ \\ 
 \hline
\end{tabularx}
\end{minipage}

\caption{
    Results from Fig.~\ref{fig:corr_3} showing the three-body correlation $\mathcal{C}^{(3\nu)}_{1iN}(t)$ for the $N=12$ neutrino system. The results computed using {\tt H1-1} were obtained using 240 shots, and those using {\tt H1-1E} with 1200 shots. The column labeled ``1-T'' is the exact value using a single Trotter step.
}
\end{table}
\begin{table}[!htb]
\centering
\begin{minipage}{0.32\columnwidth}
\centering
\begin{tabularx}{\columnwidth}{|| c | Y | Y | Y ||} 
\hline
 & \multicolumn{3}{c||}{$\mathcal{C}^{(3\nu)}_{1,2,3}(t)$} \\
 \hline
 $t$ & {\tt H1-1} & {\tt H1-1E} & 1-T\\
 \hline\hline
 $15$ & $-$ & $-0.017(20)$ & $-0.011$\\ 
 $20$ & $0.018(59)$ & $-$ & $-0.021$\\ 
 $25$ & $-$ & $0.001(25)$ & $-0.021$\\ 
 $30$ & $0.091(59)$ & $-$ & $-0.005$\\ 
 $35$ & $-$ & $0.037(26)$ & $0.004$\\ 
 $40$ & $0.024(59)$ & $-$ & $-0.016$\\ 
 \hline
\end{tabularx}
\end{minipage}
\begin{minipage}{0.32\columnwidth}
\centering
\begin{tabularx}{\columnwidth}{|| c | Y | Y | Y ||} 
\hline
 & \multicolumn{3}{c||}{$\mathcal{C}^{(3\nu)}_{2,3,4}(t)$} \\
 \hline
 $t$ & {\tt H1-1} & {\tt H1-1E} & 1-T\\
 \hline\hline
 $15$ & $-$ & $0.017(17)$ & $-0.004$\\ 
 $20$ & $-0.033(57)$ & $-$ & $-0.008$\\ 
 $25$ & $-$ & $0.023(22)$ & $-0.008$\\ 
 $30$ & $0.005(51)$ & $-$ & $-0.002$\\ 
 $35$ & $-$ & $-0.024(21)$ & $0.007$\\ 
 $40$ & $0.054(55)$ & $-$ & $0.012$\\ 
 \hline
\end{tabularx}
\end{minipage}
\begin{minipage}{0.32\columnwidth}
\centering
\begin{tabularx}{\columnwidth}{|| c | Y | Y | Y ||} 
\hline
 & \multicolumn{3}{c||}{$\mathcal{C}^{(3\nu)}_{3,4,5}(t)$} \\
 \hline
 $t$ & {\tt H1-1} & {\tt H1-1E} & 1-T\\
 \hline\hline
 $15$ & $-$ & $-0.004(16)$ & $-0.001$\\ 
 $20$ & $-0.019(53)$ & $-$ & $-0.003$\\ 
 $25$ & $-$ & $0.017(18)$ &  $-0.003$\\ 
 $30$ & $0.063(41)$ & $-$ & $-0.003$\\ 
 $35$ & $-$ & $-0.005(17)$ & $-0.004$\\ 
 $40$ & $0.101(39)$ & $-$ & $-0.014$\\ 
 \hline
\end{tabularx}
\end{minipage}
\vspace{1em}

\begin{minipage}{0.32\columnwidth}
\centering
\begin{tabularx}{\columnwidth}{|| c | Y | Y | Y ||} 
\hline
 & \multicolumn{3}{c||}{$\mathcal{C}^{(3\nu)}_{4,5,6}(t)$} \\
 \hline
 $t$ & {\tt H1-1} & {\tt H1-1E} & 1-T\\
 \hline\hline
 $15$ & $-$ & $-0.025(17)$ & $-0.0001$\\ 
 $20$ & $-0.026(49)$ & $-$ & $-0.0001$\\ 
 $25$ & $-$ & $0.008(16)$ &  $0.0005$\\ 
 $30$ & $-0.017(31)$ & $-$ & $0.001$\\ 
 $35$ & $-$ & $-0.005(12)$ & $-0.0001$\\ 
 $40$ & $0.020(27)$ & $-$ & $-0.0005$\\ 
 \hline
\end{tabularx}
\end{minipage}
\begin{minipage}{0.32\columnwidth}
\centering
\begin{tabularx}{\columnwidth}{|| c | Y | Y | Y ||} 
\hline
 & \multicolumn{3}{c||}{$\mathcal{C}^{(3\nu)}_{5,6,7}(t)$} \\
 \hline
 $t$ & {\tt H1-1} & {\tt H1-1E} & 1-T\\
 \hline\hline
 $15$ & $-$ & $-0.011(15)$ & $0.0001$\\ 
 $20$ & $-0.032(43)$ & $-$ & $0.0002$\\ 
 $25$ & $-$ & $0.003(14)$ & $0.0003$\\ 
 $30$ & $-0.022(16)$ & $-$ & $0.0003$\\ 
 $35$ & $-$ & $-0.006(11)$ & $-0.00003$\\ 
 $40$ & $-0.015(25)$ & $-$ & $-0.0002$\\ 
 \hline
\end{tabularx}
\end{minipage}
\begin{minipage}{0.32\columnwidth}
\centering
\begin{tabularx}{\columnwidth}{|| c | Y | Y | Y ||} 
\hline
 & \multicolumn{3}{c||}{$\mathcal{C}^{(3\nu)}_{6,7,8}(t)$} \\
 \hline
 $t$ & {\tt H1-1} & {\tt H1-1E} & 1-T\\
 \hline\hline
 $15$ & $-$ & $-0.002(16)$ &$-0.0001$\\  
 $20$ & $0.043(40)$ & $-$ &  $-0.0002$\\ 
 $25$ & $-$ & $0.002(14)$ &$-0.0003$\\ 
 $30$ & $-0.008(32)$ & $-$ & $-0.0003$\\ 
 $35$ & $-$ & $-0.014(13)$ &  $0.00003$\\ 
 $40$ & $0.019(13)$ & $-$ &  $0.0002$\\ 
 \hline
\end{tabularx}
\end{minipage}
\vspace{1em}

\begin{minipage}{0.32\columnwidth}
\centering
\begin{tabularx}{\columnwidth}{|| c | Y | Y | Y ||} 
\hline
 & \multicolumn{3}{c||}{$\mathcal{C}^{(3\nu)}_{7,8,9}(t)$} \\
 \hline
 $t$ & {\tt H1-1} & {\tt H1-1E} & 1-T\\
 \hline\hline
 $15$ & $-$ & $0.017(17)$ & $0.0001$\\ 
 $20$ & $0.044(43)$ & $-$ & $0.0001$\\ 
 $25$ & $-$ & $0.003(15)$ &  $-0.0005$\\ 
 $30$ & $0.001(32)$ & $-$ & $-0.001$\\ 
 $35$ & $-$ & $0.002(11)$ & $0.0001$\\ 
 $40$ & $-0.018(19)$ & $-$ & $0.0005$\\ 
 \hline
\end{tabularx}
\end{minipage}
\begin{minipage}{0.32\columnwidth}
\centering
\begin{tabularx}{\columnwidth}{|| c | Y | Y | Y ||} 
\hline
 & \multicolumn{3}{c||}{$\mathcal{C}^{(3\nu)}_{8,9,10}(t)$} \\
 \hline
 $t$ & {\tt H1-1} & {\tt H1-1E} & 1-T\\
 \hline\hline
 $15$ & $-$ & $0.021(18)$ &  $0.001$\\
 $20$ & $0.013(37)$ & $-$ &$0.003$\\ 
 $25$ & $-$ & $0.004(19)$ & $0.003$\\ 
 $30$ & $0.011(40)$ & $-$ &  $0.003$\\
 $35$ & $-$ & $0.007(16)$ &$0.004$\\ 
 $40$ & $-0.013(35)$ & $-$ &$0.014$\\ 
 \hline
\end{tabularx}
\end{minipage}
\begin{minipage}{0.32\columnwidth}
\centering
\begin{tabularx}{\columnwidth}{|| c | Y | Y | Y ||} 
\hline
 & \multicolumn{3}{c||}{$\mathcal{C}^{(3\nu)}_{9,10,11}(t)$} \\
 \hline
 $t$ & {\tt H1-1} & {\tt H1-1E} & 1-T\\
 \hline\hline
 $15$ & $-$ & $-0.011(19)$ & $0.004$\\
 $20$ & $-0.013(45)$ & $-$ &  $0.008$\\
 $25$ & $-$ & $0.042(23)$ & $0.008$\\
 $30$ & $0.007(49)$ & $-$ & $0.002$\\ 
 $35$ & $-$ & $-0.012(20)$ & $-0.007$\\ 
 $40$ & $-0.025(48)$ & $-$ & $-0.012$\\ 
 \hline
\end{tabularx}
\end{minipage}
\vspace{1em}

\begin{minipage}{0.32\columnwidth}
\centering
\begin{tabularx}{\columnwidth}{|| c | Y | Y | Y ||} 
\hline
 & \multicolumn{3}{c||}{$\mathcal{C}^{(3\nu)}_{10,11,12}(t)$} \\
 \hline
 $t$ & {\tt H1-1} & {\tt H1-1E} & 1-T\\
 \hline\hline
 $15$ & $-$ & $0.013(21)$ & $0.011$\\   
 $20$ & $0.055(53)$ & $-$ & $0.021$\\  
 $25$ & $-$ & $-0.021(25)$ & $0.021$\\  
 $30$ & $0.059(58)$ & $-$ &  $0.005$\\ 
 $35$ & $-$ & $0.026(26)$ &  $-0.004$\\ 
 $40$ & $0.001(58)$ & $-$ &  $0.016$\\ 
 \hline
\end{tabularx}
\end{minipage}

\caption{
    Results from Fig.~\ref{fig:corr_3} showing the three-body correlation $\mathcal{C}^{(3\nu)}_{i-1,i,i+1}(t)$ for the $N=12$ neutrino system. The results computed using {\tt H1-1} were obtained using 240 shots, and those using {\tt H1-1E} with 1200 shots. The column labeled ``1-T'' is the exact value using a single Trotter step.
}
\end{table}
\begin{table}[!htb]
\centering
\begin{minipage}{0.32\columnwidth}
\centering
\begin{tabularx}{\columnwidth}{|| c | Y | Y | Y ||} 
\hline
 & \multicolumn{3}{c||}{$\tau_4 (t)$} \\
 \hline
 $t$ & {\tt H1-1} & {\tt H1-1E} & 2-T\\
 \hline\hline
 $15$ & $-$ & $0.105(9)$ & $0.115$ \\ 
 $20$ & $0.088(13)$ & $-$ & $0.073$ \\ 
 $25$ & $-$ & $0.011(3)$ & $0.001$\\ 
 $30$ & $0.119(15)$ & $-$ & $0.148$ \\ 
 $35$ & $-$ & $0.406(14)$ & $0.484$ \\ 
 $40$ & $0.504(23)$ & $-$ & $0.629$\\ 
 \hline
\end{tabularx}
\end{minipage}
\begin{minipage}{0.32\columnwidth}
\centering
\begin{tabularx}{\columnwidth}{|| c | Y | Y ||} 
\hline
 & \multicolumn{2}{c||}{$\tau_6(t)$} \\
 \hline
 $t$ & {\tt H1-1E} & 1-T\\
 \hline\hline
 $17$ & $0.008(2)$ & $0.011$\\ 
 $27$ & $0.002(1)$ & $0.005$\\ 
 $37$ & $0.058(7)$ & $0.090$\\ 
 \hline
\end{tabularx}
\end{minipage}
\caption{
    Results from Fig.~\ref{fig:ntang1}, showing the $N$-tangle $\tau_N$ for the $N=4$ and $N=6$ neutrino systems. The results computed using {\tt H1-1} were obtained using 480 shots, and those using {\tt H1-1E} with 1200 shots. The column labeled ``1-T'' (``2-T'') is the exact value using a single (two) Trotter step.
}
\end{table}